\documentclass[
twocolumn,
english,
aps,
pra,
longbibliography,
superscriptaddress,
amsmath,
amssymb,
floatfix
]{revtex4-1}
\usepackage{bbold}
\usepackage{amsmath}
\usepackage{mathtools}
\usepackage{graphicx}
\usepackage{braket}
\usepackage{rotating} 
\usepackage{longtable} 
\usepackage{graphicx}
\usepackage[caption=false]{subfig}
\usepackage{multirow}
\usepackage[table,xcdraw]{xcolor}

\usepackage[pdftex,hyperfigures,pdfpagelabels,colorlinks,citecolor=blue,bookmarks=false,linkcolor=black,urlcolor=blue]{hyperref}

\usepackage{amsmath,amssymb}
\usepackage[most]{tcolorbox}
\usepackage{mathrsfs}
 \usepackage{graphicx}

\usepackage{listings}

\usepackage[utf8]{inputenc}
\usepackage{graphicx,stackrel}
\usepackage{dcolumn}
\usepackage{bm}
\usepackage[bbgreekl]{mathbbol}
\usepackage{amsmath,dsfont}
\usepackage[normalem]{ulem}
\usepackage{color}
\usepackage{silence}

\usepackage{float}
\makeatletter
\let\newfloat\newfloat@ltx
\makeatother
\usepackage{algpseudocode}

\usepackage{algorithm}

\usepackage{rotating} 

\usepackage{lipsum}
\usepackage{afterpage}

\usepackage{tikz}
\usetikzlibrary{shapes.geometric, arrows, positioning}

\algrenewcommand\algorithmicrequire{\textbf{Input:}}
\algrenewcommand\algorithmicensure{\textbf{Output:}}
\usepackage{amsmath,amssymb,amsfonts}
\usepackage[utf8]{inputenc}

\usepackage{hyperref}
\usepackage{multirow}
\usepackage{ragged2e}
\usepackage{ifthen}
\newcounter{complete}
\begin{document}
\title{Mixed Integer Linear Programming Solver Using Benders Decomposition Assisted by Neutral Atom Quantum Processor}

\author{M.Yassine Naghmouchi}
\email{yassine.naghmouchi@pasqal.com}
\affiliation{PASQAL SAS, 7 rue Léonard de Vinci, 91300 Massy, France}
\author{Wesley da Silva Coelho}
\email{wesley.coelho@pasqal.com}
\affiliation{PASQAL SAS, 7 rue Léonard de Vinci, 91300 Massy, France}

\date{\today}
\begin{abstract}
This paper presents a new hybrid classical-quantum approach to solve Mixed Integer Linear Programming (MILP) using neutral atom quantum computations. We apply Benders decomposition (BD) to segment MILPs into a master problem (MP) and a subproblem (SP), where the MP is addressed using a neutral-atom device, after being transformed into a Quadratic Unconstrained Binary Optimization (QUBO) model, \textcolor{black}{with an automatized procedure}. \textcolor{black}{Our MILP to QUBO conversion tightens the upper bounds of the involved continuous variables, positively impacting the required qubit count, and the convergence of the algorithm.}
To solve the QUBO, we develop a heuristic for atom register embedding and apply \textcolor{black}{a variational algorithm} for pulse shaping. In addition, we implement a Proof of Concept (PoC) that outperforms existing solutions. We also conduct preliminary numerical results: in a series of small MILP instances our algorithm identifies over 95\% of feasible solutions of high quality, outperforming classical BD approaches where the MP is solved using simulated annealing. To the best of our knowledge, this work is the first to utilize a neutral
atom quantum processor in developing an automated, problem-agnostic framework for solving MILPs through BD.

\end{abstract}
\maketitle

\section*{Introduction} 

Combinatorial optimization problems are crucial in industries such as logistics, planning, telecommunications, and resource management~\cite{paschos2014applications}. They offer significant economic and strategic benefits. However, as these problems increase in size, involving more variables and constraints, they become computationally challenging. Consequently, finding high-quality solutions quickly becomes a difficult task. To address these challenges, there is ongoing development in advanced classical optimization techniques. In particular, Mixed Integer Linear Programming (MILP)~\cite{benichou1971experiments} plays a crucial role in solving a wide range of optimization problems. It integrates integer and continuous variables, which adds computational complexity compared to pure Integer Linear Programming (ILP)~\cite{vanderbei2020linear}. For instance, tasks such as solution space tightening using \textit{cutting planes}, i.e., linear inequalities added to eliminate infeasible solutions, becomes a more difficult task with MILPs~\cite{marchand2002cutting}. Benders Decomposition (BD)~\cite{bnnobrs1962partitioning} is an efficient method for solving MILPs. The approach stands out for its applicability to a wide range of MILPs, unlike other structure-dependent methods such as Dantzig-Wolf~\cite{vanderbeck2006generic}. It separates integer variables in a process called \textit{restriction}~\cite{rahmaniani2017Benders}. This process splits the MILP into a Master Problem (MP) and a Linear Program (LP) subproblem (SP). While the SP is manageable on classical computers, the MP, containing discrete variables, constitutes the computational bottleneck. This work addresses this specific bottleneck using neutral
atom quantum processor in a hybrid classical-quantum framework.

In recent years, hybrid classical-quantum approaches start to gain traction in addressing NP-hard problems~\cite{farhi2014quantum, mcclean2016theory, nannicini2019performance, li2017hybrid, da2023quantum}. The approach assigns the computationally hard part, such as an ILP, to a Quantum Processing Unit (QPU). Conversely, classical Central Processing Units (CPUs) handle less computationally demanding parts like LPs. In MILP solving with BD, hybrid methods show promising results over classical methods~\cite{zhao2022hybrid, gao2022hybrid, chang2020hybrid, fan2022hybrid}. Generally, these methods uses a QPU to solve the MP after transforming it to a Quadratic Unconstrained Binary Optimization (QUBO) model. In neutral atom quantum computing, atoms, controlled by lasers, serve as qubits and are versatile enough to encode any QUBO~\cite{nguyen2023quantum}. Known for its scalability and precision, enabled by optical tweezers, this method distinguishes itself among quantum technologies such as Josephson junctions~\cite{kjaergaard2020superconducting}, trapped ions~\cite{pogorelov2021compact}, and photons~\cite{anton2019interfacing}. In this context, the Hamiltonian governing qubit dynamics can be tailored to a QUBO model in such a way that the Hamiltonian's ground state encodes the optimal solution to the QUBO~\cite{glover2018tutorial}. This elegant alignment between physics and algorithms enables to tackle a vast variety of optimization problems. The algorithm design for neutral atom systems includes \textit{register embedding} and \textit{pulse shaping}. Register embedding spatially arranges qubits to mirror the QUBO matrix, which serves as the problem's encoding method. Pulse shaping, in contrast, adjusts laser pulses to manipulate qubit states and cover strategies like the \textcolor{black}{variational algorithms~\cite{mcclean2016theory} and Quantum Approximate Optimization (inspired) Algorithms (QAOA)~\cite{farhi2014quantum}}. While the problem encoding is based on register embedding, pulse shaping directs the algorithm toward finding a solution, with each component contributing to the algorithm's overall structure.

\begin{figure}[!htb]
    \includegraphics[width=\linewidth]{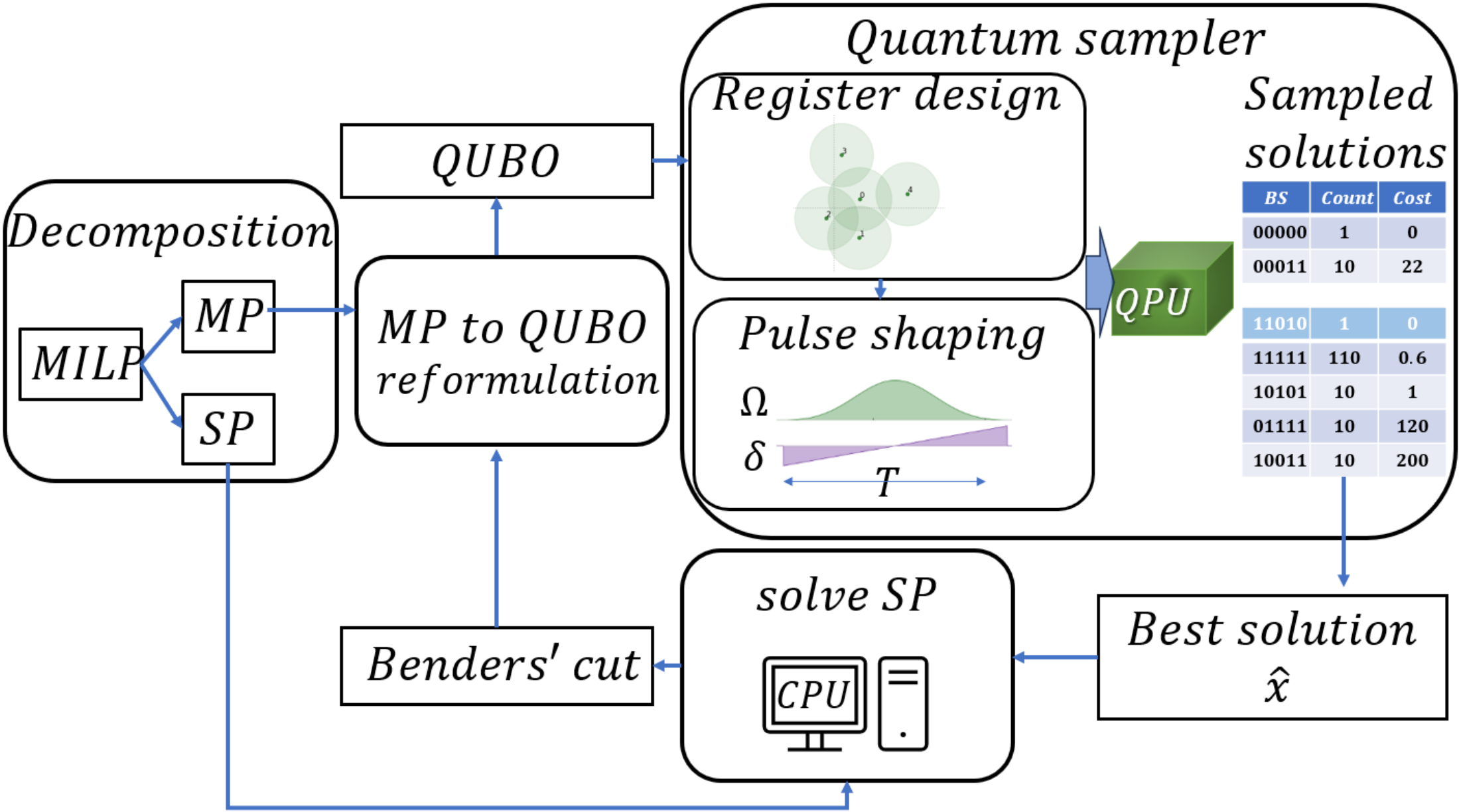}
    \caption{Approach overview.  The figure represents our hybrid framework that merges classical computing with neutral atom quantum processing for solving MILP problems. MILP is divided into a MP and a SP. The MP is reformulated into a QUBO, which is then processed by a quantum sampler. Register embedding is applied to configure qubits for QUBO encoding, and pulse shaping tunes the laser pulses, preparing the QPU for solution exploration. The best quantum-derived solution, denoted as $\hat{x}$, is assessed using the SP on a CPU which decides whether a Benders' cuts has to be added to the MP. Through iterative Benders' cuts, the MP is refined, discarding ``bad'' solutions and guiding the process toward the most effective MILP solution.}
    \label{fig:hybrid_Benders_flowchart}
\end{figure}

In this study, we propose a hybrid classical-quantum BD framework.  As presented in Figure~\ref{fig:hybrid_Benders_flowchart}, our approach begins by splitting the MILP into a MP and a SP. The MP is reformulated into a QUBO for quantum processing. Based on an iterative BD algorithm, the MP is solved using a QPU, and the SP on a CPU. Register embedding and pulse shaping algorithms are executed prior to solving the MP, yielding the creation of the Hamiltonian which enables quantum sampling of various MP binary solutions. Once the best MP solution $\hat{x}$ is fixed, the SP which uses $\hat{x}$ as a parameter is solved by classical solver like CPLEX~\cite{IBMCPLEX2023} or Gurobi~\cite{GurobiOptimization}. The SP optimal objective value is then used to assess the need for improving the current solution. If necessary, a \textit{Benders' cut}, which is a linear inequality derived from the SP, is integrated into the MP in the subsequent iteration to eliminate infeasible solutions and guide the algorithm towards optimality. Our iterative method is designed to achieve an optimal or near-optimal solution for the original MILP and is automatically adapted to various MILP scenarios.

The contributions of this study are manifold. We introduce an automated process for converting the MP into a QUBO model. First, we present a heuristic for register embedding. Then, we implement a \textcolor{black}{variational algorithm} for pulse shaping, incorporating gradient descent techniques. Additionally, we conduct a Proof of Concept (PoC) that demonstrates superior performance over a D-Wave annealer application of BD~\cite{zhao2022hybrid}. Moreover, we perform numerical results, on a large set of MILP instances. This shows our method's efficiency in comparison to a classical BD approach where the MP is solved using simulated annealing~\cite{bertsimas1993simulated}. The adaptability of BD to various MILP structures, along with the scalability and precision of neutral atoms in solving QUBOs, forms the foundation of our research motivation. To the best of our knowledge, this work is the first to utilize neutral atoms computation in developing an automated, problem-agnostic framework for solving MILPs through BD. 

This paper is organized as follows: Section~\ref{section:relatedwork} provides an overview of the developments in (hybrid) decomposition methods. Section~\ref{section:background} presents both physics and mathematical material used in this work. The conversion of the MP into a QUBO model is detailed in Section~\ref{section:to_qubo}. Section~\ref{section:quantumalgo} presents the design of the quantum algorithm. The overall hybrid algorithm is described in Section~\ref{section:hybridBenders}. Our PoC and numerical results are discussed in Section~\ref{section:experimentation}. Section~\ref{section:conclusion} concludes the paper and discusses future directions.

For a comprehensive list of acronyms used throughout this paper, the reader is refereed to Table~\ref{table:acronyms}.
\begin{table}[h]
\centering
\begin{tabular}{ll}
\hline
\textbf{Acronym} & \textbf{Definition} \\
\hline
LP & Linear Programming \\
ILP & Integer Linear Programming \\
MILP & Mixed Integer Linear Programming \\
BD & Benders Decomposition \\
OP & Original Problem \\
MP & Master Problem \\
SP & Subproblem \\
QUBO & Quadratic Unconstrained Binary Optimization \\
QPU & Quantum Processing Unit \\
CPU & Central Processing Unit \\
QAOA & Quantum Approximate Optimization Algorithm \\
PoC & Proof of Concept \\
ADMM & Alternating Direction Method of Multipliers \\
IoT & Internet of Things \\
\hline
\end{tabular}
\caption{List of Acronyms}
\label{table:acronyms}
\end{table}

\section{Related work}
\label{section:relatedwork}

Classical decomposition strategies, including methods such as Dantzig-Wolfe~\cite{vanderbeck2006generic}, and column generation~\cite{desaulniers2006column},  efficiently address NP-hard optimization problems by subdividing them into smaller, manageable parts. In recent advancements, diverse approaches are being explored to address complex problems through classical-quantum decomposition methods. An instance, as presented in~\cite{do2023quantum}, focuses on applying these methods to an Internet of Things (IoT) use-case, specifically for edge server placement and workload allocation optimization. This study employs a hybrid quantum-classical framework, partitioning the problem into a QUBO MP and an LP SP. Using the IBM quantum computing SDK Qiskit~\cite{wille2019ibm}, the authors proposes a QAOA algorithm, within the Alternating Direction Method of Multipliers (ADMM) framework~\cite{boyd2011distributed}, an optimization technique that divides complex problems into smaller, manageable SPs for iterative solving.

BD stands out among existing decomposition strategies for its applicability across a variety of MILP problems. It addresses both integer and continuous variables, while being adapted to any constraints structure. Nonetheless, classical BD encounters several challenges. Its main limitations include slow convergence to an optimal solution, and time-consuming iterations. These issues often arise due to suboptimal initial solutions, weak Benders' cuts, and the presence of multiple equivalent solutions. Addressing these issues, significant enhancements have been made to the classical BD algorithm~\cite{naoum2013interior, crainic2014partial, yang2012tighter, fabian2007solving, rubiales2013stabilization}. These include the development of advanced stabilization methods, which enhance the algorithm resilience against solution quality fluctuations, consequently accelerating convergence. Additionally, the implementation of techniques to generate multiple solutions and cuts in each iteration has been investigated and has demonstrated effectiveness in reducing the BD algorithm's time to solution~\cite{beheshti2019accelerating}.

Recently, hybrid classical-quantum BD approaches are starting to show a potential in tackling these MILPs complexities. Addressing the computational difficulty of the MP, Zhao et al.~\cite{zhao2022hybrid} develop a hybrid quantum BD approach. This method involves transforming the MILP model of the MP into a QUBO model, which is then solved using quantum annealing, on D-Wave machines. Similarly, Gao et al.~\cite{gao2022hybrid} have tackled a unit commitment problem using a hybrid quantum-classical generalized BD algorithm. This approach involves also converting the MP into a QUBO and solving it via quantum annealing on the D-Wave machine, showing improved performance over the ADMM method.

Further research by Chang et al.~\cite{chang2020hybrid} explores the hybrid BD on noisy intermediate-scale quantum processors, demonstrating efficiency in small-scale scenarios, particularly in power system-specifc MILP.  Their algorithm is tested on a D-Wave 2000Q QPU. Furthermore, Fan et al.~\cite{fan2022hybrid} investigate the application of BD in network resource optimization areas like network function virtualization and multi-access edge computing. Franco et al.~\cite{franco2023efficient} demonstrate that while BD applies to a wider range of MILP problems, potentially requiring more qubits in extreme cases, Dantzig-Wolfe is more qubit-efficient but limited to structurally constrained problems. On the quantum side, the authors use the D-Wave system. 

While significant progress has been made in hybrid BD using annealing and gate-based quantum computing methods, to the best of our knowledge, the exploration of neutral atom quantum computing within this framework remains an unexplored area, which is the purpose of our current research.

\section{Background}
\label{section:background}

This section presents the technical tools used in our research, including a brief introduction to quantum computing with neutral atoms, and the foundations of classical BD. 

\subsection{Neutral Atoms QPUs \textcolor{black}{and Optimization Problems}}

Neutral atom QPUs utilize state transitions in valence electrons of atoms like Rubidium to establish qubit states. These states include the stable ground state (\(\ket{g}\) or \(\ket{0}\)) and an energized Rydberg state (\(\ket{r}\) or \(\ket{1}\)). In these QPUs, the atoms arranged in a specific spatial configuration, the so-called register, contribute in defining an effective Hamiltonian \(H(t)\), which governs the dynamics of the quantum system. The Hamiltonian is given by:
\begin{equation}
\label{hamiltonian}
    H(t) = \Omega(t)\sum_{u=1}^{|V|}\hat{\sigma}^x_u - \Delta(t)\sum_{u=1}^{|V|}\hat{n}_u + \sum_{u<v=1}^{|V|} U_{uv} \hat{n}_u \hat{n}_v. 
\end{equation}
Here, \(\Omega(t)\) denotes the Rabi frequency amplitude that controls the rate of state transitions of qubits, \(\Delta(t)\) represents the laser detuning which affects the probability of atom excitation, and \(U_{uv}\) is the interaction strength between atoms \(u\) and \(v\). The terms \(\hat{\sigma}^x_u\) and \(\hat{n}_u\) correspond respectively to the Pauli X operator, and \(\hat{n}_u = (\mathbb{I} +  \hat{\sigma}^z_u)/2\). Finally, \( V \) represents the set of atoms in the register.

\color{black}

Neutral atom QPUs naturally solve the Maximum Independent Set (MIS) problem~\cite{tarjan1977finding} in unit disk graphs~\cite{clark1990unit}. The MIS problem, which involves finding the largest set of non-adjacent vertices (atoms), is optimally addressed by these QPUs, thanks to the Rydberg blockade mechanism. This mechanism uses the principle that two atoms within a certain distance cannot both be in an excited state simultaneously. In the context of encoding unit disk graphs, an edge between two vertices exists if they are close enough for the Rydberg blockade to take effect—essentially, if the atoms are within the blockade radius, they are considered adjacent in the graph. This spatial encoding reflects the edges and vertices of the unit disk graph directly into the quantum system, facilitating a native solution of the MIS. 

More generally, neutral atom QPUs are capable of solving any optimization problems that can be encoded into a QUBO format, as demonstrated by studies like~\cite{nguyen2023quantum, stastny2023functional}. Here, the Rydberg blockade mechanism offers a method for the encoding of optimization problems into a QUBO format. Each pair of adjacent atoms, within the blockade radius, can be seen as a product of two binary variables multiplied by the interaction of the atoms. Solving a QUBO problem with neutral atom devices entails two major phases:
\subsubsection{Problem encoding via register embedding}
In this initial phase, called \textit{register embedding}, the problem is embedded into the QPU register by spatially arranging the atoms. The spatial configuration of the atoms is designed a in such way that their interactions, \( U_{uv} \) in equation~\eqref{hamiltonian}, represent the off-diagonal terms of the QUBO instance. This setup is crucial as the quality of the solution for the input QUBO instance is highly dependent of how well the resulting interactions $U_{uv}$ mimic the off-diagonal QUBO coefficients.

\subsubsection{Problem solving via pulse shaping}

The second phase includes the design and execution of specific pulse sequences to guide the system towards a good solution to the original QUBO. Pulse design, a process known as \textit{pulse shaping}, is essential for finalizing the Hamiltonian construction. It involves setting parameters such as pulse duration \(T\), laser detuning \(\Delta(t)\), and  the Rabi frequency \(\Omega(t)\). The choice and the effectiveness of pulse shaping in solving the problem using a neutral atoms QPU depends on the encoding phase:
        \begin{itemize}
            \item \textit{Pulse shaping under exact encoding}: If the problem is encoded exactly during the encoding phase, adiabatic pulses can be designed. These pulses leverage the adiabatic theorem~\cite{Sakurai_ModernQM} to ensure the system smoothly transitions toward its ground state, which is ideal for problems like the MIS in unit disk graphs~\cite{tarjan1977finding}. In such cases, careful adjustments of \(\Omega(t)\) and \(\Delta(t)\) over time guide the system adiabatically from its initial state to the desired final state, maintaining the system in its ground state throughout and leading to high quality solutions.
            \item \textit{Pulse shaping under approximate encoding}: When exact encoding is not feasible, variational algorithms are employed to design pulses that still aim to steer the system toward effective solutions. Although these pulses may not be adiabatic, they are crafted to achieve satisfactory results by approximating the desired Hamiltonian dynamics, thus navigating the system towards good solutions for the input QUBO instance under constrained or imperfect encoding conditions.
            \end{itemize}

Note that neutral atom QPUs are resilient to noise, which helps in the efficient solving of combinatorial optimization problems. As highlighted in~\cite{da2022efficient}, errors in digital quantum computing can propagate from one gate to another, potentially compounding as the computation progresses. In contrast, the analog evolution in neutral atom quantum computing inherently reduces error propagation. This robustness is due to the direct correspondence between the state of the computational basis in which the qubits are measured and the solution to the optimization problem. In some cases, noise can even enhance performance by helping the system explore a broader solution space, as suggested in~\cite{novo2018environment}. 



\color{black}

For a in-depth understanding of neutral atom-based quantum computing, one may refer to~\cite{henriet2020quantum}.

\subsection{Benders Decomposition for MILPs}
\label{section:Bendersdecompo}

BD works by separating a MILP known as the Original Problem (OP) into a MP, which deals with discrete variables, and a SP, which focuses on the remaining variables, often continuous. To perform BD, the OP is reformulated into an equivalent MILP. This reformulation relays on the principles of polyhedral and  duality theory in linear optimization~\cite{Benders2005partitioning}. Consider real matrices \( A \)  of dimensions \( m_1 \times n \), \( G \) of dimensions \( m_1 \times p \), and \( B \) of dimensions \( m_2 \times n \). Let \( c \), \( h \), \( b \) and \( b' \)  be vectors with dimensions \( n \), \( p \), \( m_1 \), and \( m_2\), respectively. The OP is initially expressed as:
\begin{align}
\label{OP} \max_{x,y} \quad & c^T x + h^T y \\
\label{linking-constraints} \text{s.t.} \quad & Ax + Gy \leq b, \\
\label{x-constraints} & Bx \leq b', \\
\label{x-variables} & x \in \{0, 1\}^n, \\
\label{y-variables} & y \in \mathbb{R}_+^{p}.
\end{align}

Constraints \eqref{x-variables} are integarlity constraints, they introduce binary decision variables \(x\). Constraints \eqref{y-variables} introduce non-negative continuous decision variables \(y\). The objective function \eqref{OP} consists in maximizing a linear function of $x$ and $y$. The constraints \eqref{linking-constraints} associate the binary variables to the continuous ones, while \eqref{x-constraints} exclusively involves the binary variables. 

Following the principles of standard BD, constraints \eqref{linking-constraints} are included in the SP, with the objective function \(h^T y\). On the other hand, the constraints \eqref{x-constraints}, are incorporated into the \textcolor{black}{MP}.

For a fixed solution \(\hat{x}\) from the \textcolor{black}{MP}, the LP version of the SP is:
\begin{align}
\label{sp:objective} \max_{y} \quad & h^T y \\
\label{sp:constraints}  \text{s.t.} \quad & A\hat{x} + Gy \leq b, \\
\label{sp:variables}  & y \in \mathbb{R}_+^{p}.
\end{align}

Its dual, denoted as SP-D is:
\begin{align}
\min_{\mu} \quad & f(\hat{x}) = (b - A\hat{x})^T \mu \\
\text{s.t} \quad & G^T \mu \geq h,\\
& \mu \in \mathbb{R}^m_+.
\end{align}

Let $x^*$ be the optimal solution of SP, and $y^*$ be the optimal solution of SP-D. By strong duality theory~\cite{balinski1969duality}, the optimal objective value of the SP is equal to the optimal objective value of SP-D. We have that $$f(x^*) = h^T y^*.$$

In linear programming, the set of all feasible solutions defined by linear constraints forms a polyhedron. For a bounded polyhedron, the vertices, known as \textit{extreme points}, are key to finding potential optimal solutions. According to the corner-point theorem, also known as fundamental theorem of linear programming~\cite{schrijver1998theory}, if an optimal solution exists in a bounded linear program, it will be located at one of these vertices. In contrast, unbounded linear programs may still have extreme points, but they are not guaranteed to provide bounded optimal solutions. Instead, these unbounded problems can exhibit directions along which the objective function can increase indefinitely without violating the constraints. These directions are characterized by vectors known as \textit{extreme rays}, emerging from the polyhedron and indicating where the objective function can grow indefinitely while still satisfying all the constraints.

The feasible solution space of an LP problem is well known to be characterized by a combination of its extreme points and extreme rays, as established by Minkowski's Theorem~\cite{bertsimas1997introduction}. In the context of BD, applying this theorem to SP-D enables the reformulation of the OP into an equivalent MILP model as the following:
\begin{align}
\label{master-problem} \max_{x,\Phi} \quad & c^{T}x + \phi \\
\label{optimality-cut}\text{s.t.} \quad & (b - Ax)^{T}\mu_{o} \geq \phi, & \forall o \in \mathcal{O}, \\
\label{feasibility-cut} & (b - Ax)^{T}r_{f} \geq 0, & \forall f \in \mathcal{F}, \\
\label{x-cut} & Bx \leq b', \\
\label{master:x} & x \in \{0, 1\}^{n}, \\
\label{master:y} & \phi \in \mathbb{R}.
\end{align}

In this reformulated OP, only one continious variable \(\phi\) is used. The remaining variables are the binary variables \(x\). Constraints \eqref{optimality-cut} and \eqref{feasibility-cut} represent the Benders' \textit{optimality cuts} and \textit{feasibility cuts}, respectively. Here, the sets
\(\mathcal{O}\) and \(\mathcal{F}\) are the extreme points and extreme rays  of SP-D, respectively, and the vectors \(\mu_o\) for extreme points and \(r_f\) for extreme rays are used to construct \textit{Benders' cuts}. They can be obtained through conventional solvers, such as ILOG CPLEX~\cite{IBMCPLEX2023}. Note that sets \(\mathcal{O}\) and \(\mathcal{F}\) can be exponential in number. However, BD often efficiently finds an optimal solution using a selected subset of these sets. 

\subsection{Principle of Benders Decomposition Algorithm}

The approach begins with a restricted version of the MP \eqref{master-problem} - \eqref{master:y}, initially setting both optimality and feasibility cut sets, \(\mathcal{O}\) and \(\mathcal{F}\), to empty (\(\mathcal{O}= \mathcal{F} = \emptyset\)). As the algorithm progresses, it iteratively adds optimality and feasibility cuts to the MP. At each iteration, the MP is solved to derive an optimal solution, which is then used as a parameter to solve the SP. 

If the SP is feasible, an optimality cut is generated, guiding future MP solutions towards the OP's optimal solution. This process uses the real variable \(\phi\) in the formulation~\eqref{master-problem} - \eqref{master:y}. The optimal objective value of the SP, $(b - Ax)^{T}\mu_{o}$ is compared to~\(\phi\), and if~$(b - Ax)^{T}\mu_{o} \leq \phi$, the optimality cut \eqref{optimality-cut}, constraining \(\phi\) to not exceed the objective value of the actual extreme point of SP-D, is added. On the other hand, as the OP maximizes~\(\phi\), at optimality, \(\phi\) will be necessarily equal the objective value to SP-D, and thereby equal the optimal objective value of SP (according to the strong duality theorem). Conversely, if the SP is infeasible, a feasibility cut is produced to eliminate infeasible solutions from the MP’s solution space. These cuts act as a filter, maintaining consistency within the solutions of the~OP. 

The algorithm operates as follows. The MP try to maximize the value of \(\phi\), yielding an upper bound of the latter, while the SP, upon finding a feasible solution, generates an optimality cut that decreases this upper bound. This iterative process of maximizing and constraining \(\phi\) continues until the optimality cut from the SP can no longer decrease the value of \(\phi\). At this point, we have that $(b - Ax)^{T}\mu_{o} \geq \phi$,  and the algorithm terminates, indicating that the MP and SP have converged to an optimal solution for the OP. \textcolor{black}{Algorithm~\ref{alg:classical_Benders_decomposition} outlines the implementation of the classical Benders' decomposition method.}

\begin{algorithm}
\color{black}
\caption{Classical Benders' Decomposition}
\label{alg:classical_Benders_decomposition}
\begin{algorithmic}[1]
\Require Problem parameters \(c, h, A, G, b, B, b' \) 
\Require \( \phi_{\text{max}} \) \Comment{Initial upper bound for \( \phi \)}
\State Initialize \( \mathcal{O} \gets \emptyset, \mathcal{F} \gets \emptyset \)
\State \( \phi \gets \phi_{\text{max}} \)
\State convergence \( \gets \) False
\While{not convergence}
    \State Solve the MP with current cuts 
    \State \( x, \phi \gets \) solution of MP
    \State Solve the SP using \( x \) to find \( y \)
    \If{SP is feasible}
        \State \( \mu_o \gets \) dual variables from SP
        \State \( cut\_value \gets (b - Ax)^T \mu_o \)
        \If{\( cut\_value < \phi \)}
            \State Add optimality cut to \( \mathcal{O} \): \( \phi \leq cut\_value \)
        \Else
            \State convergence \( \gets \) True
        \EndIf
    \Else
        \State Generate a feasibility cut based on SP infeasibility
        \State Add this cut to \( \mathcal{F} \)
    \EndIf
\EndWhile
\State \Return Optimal solution \( x, \phi, y\)
\end{algorithmic}
\end{algorithm}

\color{black}

It is important to note that the resolution of the MP presents considerable complexity, mainly attributed to the fact that the latter integrates binary variables. This complexity becomes more critical as the solution process progresses, particularly with the the dynamic integration of Benders' cuts. Consequently, the MP frequently becomes a computational bottleneck. To overcome this, we study the application of  neutral atom-based quantum computing in a complete hybrid classical-quantum framework. In the next section, we detail the reformulation of the MP into a QUBO model, which is suited for this type of QPUs.

\section{Master problem reformulation to QUBO}
\label{section:to_qubo}

A Quadratic Unconstrained Binary Optimization (QUBO) problem can be formulated as follows:
\begin{equation}
\nonumber \min_z \{z^T Q z \,|\, z \in \{0, 1\}^t \}, 
\end{equation}
where \( Q \) is a symmetric matrix and \( z \) is a binary vector. The objective of this optimization problem is to find the binary vector \( z^* \) that minimizes the quadratic objective function \( z^T Q z \) over all binary vectors. Before using a QPU for optimizing the MP, we transform the latter, originally represented as a MILP, into a QUBO model. In what follows we detail the methodology of this reformulation, associated to a single BD iteration.  \textcolor{black}{For foundational insights into the MILP to QUBO conversion, refer to ~\cite{glover2018tutorial}. Note that similar methodologies are discussed in ~\cite{zhao2022hybrid}. Here, we give a detailed approach that further refines the upper bounds of the continuous variables resulting from the conversion, which positively impacts the required number of qubits as well as the convergence of the algorithm.}

\subsection{Master problem objective function reformulation}

The objective function in the MP, as defined in \eqref{master-problem}, contains a linear term and a continuous variable, given by \( c^T x + \phi \). The linear component \( c^T x \), which exclusively involves binary variables, is directly adaptable to a QUBO. This is achieved by employing the diagonal matrix \(\text{diag}(c)\), with the vector \( c \) populating its diagonal, thereby transforming \( c^T x \) into 
\begin{equation}
\label{obj-x} H_{c} = x^T \text{diag}(c) x.
\end{equation}

The continuous variable \( \phi \) can be binary encoded using a binary vector \( w \) of length \( L \), formulated as:
\begin{equation}
H_{\phi} = \sum_{i=0}^{P-1} 2^i w_i + \sum_{j=1}^{D} 2^{-j} w_{P + j} - \sum_{k=1}^{N} 2^{k-1} w_{P + D + k}.
\end{equation}

Here, \( P \) represents the number of bits for the positive integer part of \( \phi \). We have that \( P = \lfloor \log_2(\phi_{\text{max}}) \rfloor + 1\), where \( \phi_{\text{max}} \) is an upper bound of \( \phi \). For the fractional part, the number of bits \( D \) can be obtained  based on a desired precision \( \epsilon \), calculated as \( D = \lfloor \log_2(\epsilon) \rfloor + 1 \). \( N \) denotes the number of bits for the negative integer part of \( \phi \), leading to a total length \( L = P + D + N \) for the vector \( w \). Note that careful determination of \( P \), \( D \), and \( N \) is crucial, as it affects both the numerical precision of \( \phi \) and the quantum resource requirements in terms of the number of qubits needed. 

\color{black}
To determine \( \phi_{\text{max}} \), we adress \textit{the linear relaxation} of the formulation of the OP as given by \eqref{OP} -  \eqref{y-variables}, without considering $c^T x$ in the objective function. The linear relaxation is the formulation obtained by relaxing integer constraints \eqref{x-variables}, allowing continuous values, \textcolor{black}{and given by}:
\color{black}
\begin{align}
\label{OP-LR} \max_{x,y} \quad & \phi_{\text{max}} = h^T y \\
\label{linking-constraints-LR} \text{s.t.} \quad & Ax + Gy \leq b, \\
\label{x-constraints-LR} & Bx \leq b', \\
\label{x-variables-LR} & x \in [0, 1]^n, \\
\label{y-variables-LR} & y \in \mathbb{R}_+^{p}.
\end{align}

\color{black}

\subsection{Constraints reformulation}

\paragraph{Master Constraints:}
The reformulation of the MP constraints \eqref{x-constraints} involves integrating \textit{slack variables}, a common technique in classical optimization for transforming inequalities into equalities. The inequalities \( Bx \leq b' \) are first converted into the equalities \( Bx + s_m - b' = 0 \), where \( s_m \) is a vector of continuous positive variables. Let $1 \leq k \leq m_2$. The $k^{th}$ component, $s_m^k$, of the slack vector \( s_m\) undergoes a binary encoding yielding:
\begin{equation}
\nonumber s_m^k = \sum_{i=0}^{Q_1^k-1} 2^i v_{i}^{m, k} + \sum_{j=1}^{R_1^k} 2^{-j} v^{m, k}_{Q_1 + j}. 
\end{equation}

Here, \( v^{m,k} \) denotes a binary vector with a length of \( Q_1^k + R_1^k\), where \( Q_1^k \) represents the number of qubits required for the integer part of \( s_m^k \), and \( R_1^k \) corresponds to the number of qubits for its fractional part. The value of \( Q_1^k \) can be determined by \( \lfloor \log_2(s_{\text{max}}^k) \rfloor + 1 \), where \( s_{\text{max}}^k \) is the upper bound for \( s_m^k\) obtained by solving the following linear program
%
\color{black}
\begin{align}
\label{OP-LRs}  s_{\text{max}}^k=  \max_{x,y} \quad & b'_k - B_kx  \\
\label{linking-constraints-LRs} \text{s.t.} \quad & Ax + Gy \leq b, \\
\label{x-constraints-LRs} & Bx \leq b', \\
\label{x-variables-LRs} & x \in [0, 1]^n, \\
\label{y-variables-LRs} & y \in \mathbb{R}_+^{p}.
\end{align}

\color{black}

Let \(\pi_1^k\) be a positive penalty coefficient associated to the $k^{th}$ MP constraint. The QUBO reformulation of constraints \eqref{x-constraints} is thus given by
\begin{equation}
\label{master-qubo-constraints} H_M =  \sum\limits_{k=1}^{m_1}\pi_1^k (B_kx + s_m^k - b_k)^2,
\end{equation}

Minimizing the Hamiltonian $H_M$ to zero ensures that the constraints \( Bx \leq b' \) are satisfied. Otherwise, a cost on any deviation from zero, unsatisfying the the constraint, is added. 

\paragraph{Optimality and feasibility cuts:}
Let $C$ be the number of Benders' cuts added during the process, and let $1 \leq k \leq C$ denotes the $k^{th}$ Benders' cut. We denote by \( v^{o, k}\) and \( v^{f, k}\), respectively, the binary vectors of lengths \( Q_2^k + R_2^k\) and \( Q_3^k+ R_3^k\) used in reformulating, respectively, the $k^{th}$ optimally and feasibility cut \eqref{optimality-cut} or  \eqref{feasibility-cut}, depending on the type of the cut added in the $k^{th}$ iteration. Here, \( Q_2^k \) and \( Q_3^k \) denote the number of qubits allocated for the integer parts of slack variables \( s_o^k\) and \( s_f^k\). The terms \( R_2^k\) and \( R_3^k\) correspond to the number of qubits used for the fractional part of \( s_o^k\) and \( s_f^k\), respectively. These can be obtained based on a desired precision. 

Following the same steps used for the MP constraints, we obtain:
\begin{equation}
\label{slack2}
 s_{o}^k = \sum_{i=0}^{Q_2^k-1} 2^i v^i_{o,k} + \sum_{j=1}^{R_2^k} 2^{-j} v^{o,k}_{Q_2 + j}, \text{ and,}
\end{equation}
\begin{equation}
\label{slack3}
  s_f^k = \sum_{i=0}^{Q_3^k-1} 2^i v^{f,k}_{i} + \sum_{j=1}^{R_3^k} 2^{-j} v^{f,k}_{Q_3 + j}. 
\end{equation}

Let $\Pi_2^k$ be a positive penalty coefficient associated to the optimality cut, and $\Pi_3^k$ the one associated to the feasibility cuts. Using \eqref{slack2} and \eqref{slack3}, we obtain the Hamiltonians
\begin{equation}
\label{optimality-penalty} H_O = \pi_2^k \left(H_\phi + (\mu_o^k)^T A x +  s_o^k - b^T \mu_o^k\right)^2, \text{ and}
\end{equation}
\begin{equation}
\label{feasibility-penalty} H_F = \pi_3^k \left((r_f^k)^T A x +  s_f^k - b^T r_f\right)^2. 
\end{equation}

Here, $\mu_o^k$ represents the dual value and $r_f^k$ denotes the dual ray, both of which are obtained by solving the \(k^{th}\) SP. Note that each SP has the potential to yield either a dual value, in cases where it is feasible, or an extreme ray, if it is infeasible. Consequently, for any given SP, only one of the Hamiltonians,  \eqref{optimality-penalty} and \eqref{feasibility-penalty}, is applicable. 

The QUBO formulation of the OP is given by the sum of the Hamiltonians:
\begin{equation}
\label{QUBO} H_P=H_{\phi} + H_c + H_M +  H_O + H_F.
\end{equation}

\subsection{Reformulation discussion}
In the process of converting the MP into the QUBO model, several algorithmic challenges arises.  

\paragraph{Qubit count limitation \textcolor{black}{and convergence of the algorithm}}
The process of quadratizing the variable \( \phi \) and the constraints in the BD \eqref{master-problem} - \eqref{x-cut}, necessitates the use of additional qubits. The total number of qubits, of the initial MP, is given by 
\[ t = n + P + D + N + \left(\sum\limits_{i=1}^{m_2} Q_1^i + R_1^i\right). \]
As the algorithm progresses, this number can significantly increase, primarily due to the generation of Benders' cuts. Each iteration $k$ increases the number of qubits by $Q_2^k + R_2^k$, or $Q_3^k + R_3^k$ depending on the type of the cut. A critical consideration in this process is the current limitations of quantum computing hardware, particularly in terms of qubit availability. To ensure that the computation remains feasible on existing quantum computers, it is crucial to accurately estimate the upper bounds of the real variable \( \phi \) as well as slack variables \( s \). These estimations directly impact the number of qubits required. As previously established, we utilize the linear relaxation of the OP to calculate upper bounds. 

\color{black}
Addressing the linear relaxations defined in equations \eqref{OP-LR} to \eqref{y-variables-LR} and \eqref{OP-LRs} to \eqref{y-variables-LRs} tightens the upper bounds of \(\phi_{\text{max}}\) and the slack variables \(s_{\text{max}}^k\) compared to the method described by Zhao et al.~\cite{zhao2022hybrid}. This improvement results from incorporating additional valid constraints \eqref{linking-constraints-LR} and \eqref{x-constraints-LR} into the linear programs, which also leads to tighter values for \(\phi_{\text{max}}\) and \(s_{\text{max}}^k\).

By providing tighter upper bounds, the algorithm reaches the optimal solution more quickly as the bounds limit the solution space the algorithm needs to explore, thus accelerating convergence. Additionally, solving these linear continuous programs is generally computationally easy using classical methods.

\color{black}

\paragraph{Qubit count vs numerical precision}
An accurate estimation of qubit count is essential to avoid numerical precision issues. In fact, a bad estimation of this count can lead to significant numerical precision issues. For instance, an underestimation of $\phi$ could stop the algorithm before its effective end, resulting in poor quality solutions. Conversely, an overestimation of $\phi$ can cause almost endless loops, or the generation of no good Benders' cuts, thus resulting in unfeasible solutions. Therefore, it is important to find a balance between precision and qubits availability. 

\paragraph{Penalty values and solution efficiency}
Finally, the selection and tuning of penalty values in the QUBO model are also important. These values guide the quantum algorithm towards optimal or near-optimal solutions. \textcolor{black}{However, the process of setting static penalty weights for various types of problems is not trivial. This is because values that are too small will lead to infeasible solutions while values that are too large may lead to slower convergence. Many studies, are exploring different methods of setting penalty weights within the context of QUBO formulations~\cite{ayodele2022penalty, verma2022penalty, garcia2022exact}. The study of the best penalty configuration is not within the scope of this work}
Finding the optimal penalties for the QUBO is still being an open research topic. 

Future directions addressing the presented challenges will be discussed in the conclusion of this paper (see Section~\ref{section:conclusion}). 
The next section presents the methodology to solve the QUBO in a neutral atoms QPU. 
\section{Quantum algorithm design}
\label{section:quantumalgo}

The development of a quantum algorithm in a computer based on neutral atoms includes two major steps: register embedding and pulse shaping. In this section, we present the algorithms developed for these steps. 

\subsection{QUBO embedding strategy}
\label{subsection:embedding}

The register embedding involves placing atoms at specific locations within a register having its own distance constraints. The aim is to find the placement aligning the interaction matrix $U$, created based on the distances between the placed atoms and a device-specific constant, as closely as possible to the predefined QUBO matrix. 

Formally, the problem can be defined as the following. Given:
\begin{itemize}
    \item A register defined by a set of positions $P$, respecting a minimum distance between each pair of positions, and a maximum distance of any position from a reference point $c$, refereed as the center of the register;
    \item a set $V$ of $n$ atoms to be placed inside the register. Let $\mathcal{P}$ be the set of injective mappings from $V$ to $P$. Each mapping $\Pi : V \rightarrow P$ in $\mathcal{P}$ associates a position $p \in P$  for each atom $v \in V$. The placement of atoms on a subset of positions yields an interaction matrix $U_\Pi$, whose components are defined by $u_{ij}=\frac{C_6}{r_{\Pi(i),\Pi(j)}^6}$\textcolor{black}{~\cite{henriet2020quantum}}. Here, $C_6$ is a device-dependent constant, and $r_{\Pi(i),\Pi(j)}$ denotes the distance between the atoms $i, j \in V$ induced by the placement $\Pi$;
    \item $n$-dimensional QUBO matrix $Q$.
\end{itemize}

The register embedding problem (REP) consists in selecting the best placement $\Pi^* \in \mathcal{P}$, that minimizes the distance between the interaction matrix $U$ and the QUBO matrix $Q$:
\color{black}
\begin{equation}
\label{eq:distance}
\min_{\Pi} \sum\limits_{(i, j) \in V^2: i \neq j }|q_{\Pi(i),\Pi(j)} -  u_{\Pi(i),\Pi(j)}|.
\end{equation}
\color{black}
The REP is an NP-Hard problem. To address this complexity, we develop the heuristic presented in Algorithm~\ref{alg:register_embedding}. The algorithm starts with a random selection of an atom from the set \( V \). The selected atom is placed at the center of the register \( c \), and the process continues by iteratively evaluating and embedding the remaining atoms. 

At each iteration, the algorithm examines all available positions for the selected atom. It computes the total deviation from the desired QUBO matrix. The position that yields the lowest deviation is chosen as the best placement for that atom. For each atom \( u \) in \( V \), we compute, the position of the lowest deviation is the one that minimizes the sum of absolute value differences between the elements of the QUBO matrix \( Q \) and the interaction matrix \( U \).  The interaction matrix \( U \) is dynamically updated as each atom is placed, reflecting the current state of interactions in the register.

Once a position is selected, it is removed from the set \( P \) of available positions, and the atom is removed from the set \( V \) of unplaced atoms. The process is repeated until all atoms are placed, resulting in a configuration that finds a solution of \eqref{eq:distance}, minimizing the distance between the interaction and QUBO matrices.

\begin{algorithm}
\caption{Register Embedding Algorithm}
\label{alg:register_embedding}
\begin{algorithmic}[1]
\Require $V$ \Comment{Set of atoms.}
\Require $P$ \Comment{Set of positions.}
\Require $Q$ \Comment{QUBO matrix.}
\State Randomly select an atom $u$ from $V$ 
\State $P_a \gets (u, c)$ \Comment{ \textcolor{black}{$P_a$} is the set of atom-position pairs representing the placement of atoms. Initially contains \textcolor{black}{$(u,c)$ which represents the placement of the randomly chosen atom $u$ on the center of the register $c$}.}
\State $U \gets \text{0}_{\mathbb{R}_{n \times n}}$ \Comment{initially  no interactions}
\State \textcolor{black}{$P \gets P \setminus \{c\}$}
\Comment{\textcolor{black}{Remove $c$ from $P$}}
\State$V \gets V \setminus \{u\}$  \Comment{Remove atom $u$, already embedded}
\While{$V \neq \emptyset$}
    \State Select atom $u$ from $V$
    \State Initialize $\text{min\_sum} \gets \infty$
\State For each atom $p$ in $P$:
\State \ \ \ \  $sum \gets 0$
\State \ \ \ \ For each position $v$ such that there exists $(v,p_v) \in P_a$:
\State \ \ \ \ \ \ \ \ $sum \gets sum + |Q_{u,v} - U_{p,\Pi(v)}|$ 
\color{black}
\State \ \ \ \  If $sum < \text{min\_sum}$ then:
\State \ \ \ \ \ \ \ \ $\text{min\_sum} \gets sum$
\State \ \ \ \ \ \ \ \ $\text{best\_position} \gets p$
\color{black}
\State  \ \  $P_a \gets P_a \cup \{(u,{\text{best\_position}\})}\}$ \Comment{Place $u$ on $p$}
\State \ \ $P \gets P \setminus \{\text{best\_position}\}$   \Comment{Position $\text{best\_position}$ no longer available}
\State \ \  $V \gets V \setminus \{u\}$  \Comment{Remove atom $u$, already embedded}
\EndWhile
\State \Return $P$
\end{algorithmic}
\end{algorithm}

\color{black}
It is worth noting that deviations between \( U \) and \( Q \) are critical because they can lead to suboptimal or even infeasible solutions for the MP. Such outcomes can result in weak or no good Benders' cuts. Specifically, weak cuts are those that do not significantly tighten the value of \(\phi\), thereby failing to accelerate the convergence towards the best solution. No good cuts are invalid for the original problem (OP), potentially rendering the OP itself infeasible. These aspects highlight the importance of achieving an accurate register embedding method, as they directly impact the quality and feasibility of the solution to the OP.

\subsection{Variational Algorithm for Pulse Shaping}

This section presents a variational algorithm for optimizing pulse parameters. Our objective is to identify the optimal settings for pulse parameters for a register embedded using the heuristic detailed in Section~\ref{subsection:embedding}: the maximum amplitude \(\Omega_{\text{max}}\), the initial detuning \(\delta_{\text{init}}\), the final detuning \(\delta_{\text{final}}\), and the pulse duration \(T\), all within predetermined bounds. Since this procedure aims to find the optimal shape for the laser pulses that control the quantum system, it will be referred to as \textit{pulse shaping}. 

An iterative optimization procedure is established, progressively refining these parameters to enhance the pulse's efficiency. Initially, the average value of the parameters is used to construct a pulse, with a chosen shape (such as an interpolated waveform). This pulse is then executed on the register, which is initialized in the Rydberg state,
\(
|\psi_0\rangle.
\) The evolution of the quantum state under the influence of the pulse is described by an effective Hamiltonian \(H_{\text{eff}}\), as expressed in Eq.~\eqref{hamiltonian}. This effective Hamiltonian governs the dynamics of the system, and is not necessarily unitary. The final state \(|\psi_f(\delta_{\text{init}}, \delta_{\text{final}}, T)\rangle\) after applying the pulse, during a time $T$, is given by:
\[
|\psi_f(\delta_{\text{init}}, \delta_{\text{final}}, T)\rangle = H_{\text{eff}} |\psi_0\rangle.
\]

If the system consists of \(M\) atoms, the final state will typically be a normalized superposition of basis states, each uniquely corresponding to a binary bitstring of length \(M\): 
\[
|\psi_f\rangle = \sum_{i=1}^{2^M} a_i |b_i\rangle ,
\] where \(\sum_i |a_i|^2 = 1\), and
\[
|b_i\rangle = |b_i^1\rangle \otimes \ldots \otimes |b_i^M\rangle, \quad b_i^j = |0\rangle \text{ or } |1\rangle.
\]

Achieving perfect knowledge of the quantum state \(|\psi_f\rangle\) (and thus the coefficients \(a_i\)) would require an exponential amount of resources as the system scales. Therefore, the state is usually only approximately known through repeated measurements. Each measurement of the state \(|\psi_f\rangle\) results in the extraction of one of the bitstrings \(b_i\) with probability \(|a_i|^2\). Collecting \(N\) samples of the state yields a set of pairs \(
\{(b_i, w_i^{(N)})\}_{i=1,\ldots,2^M}
\), where \(w_i^{(N)}\) denotes the number of times the bitstring \(b_i\) was measured out of \(N\) tries. 

To each bitstring \(b_i\), a cost \(C(b_i) = H_P(b_i)\) can be assigned according to the objective value of the QUBO \(H_P\)~\eqref{QUBO} for \(b_i\). The effectiveness of the optimization is evaluated by calculating the expectation value of the problem Hamiltonian (\(H_P\)) over all samples. Specifically, the average cost from all the samples is computed as:
\[
\langle C \rangle = \frac{1}{N} \sum_i w_i C(b_i).
\]

Gradient Boosted Regression Trees (GBRT)~\cite{prettenhofer2014gradient}, a refined numerical optimization technique, are then applied to identify the parameter set that minimizes the cost function, thereby finding the optimal parameters that achieve the minimum average cost. This sampling and optimization cycle is executed repeatedly, a total of \(p\) times. The parameters and sample collection that yield the most favorable cost will be selected. The choice of \(p\) is important, as it affects both the solution's quality and the overall cost of the iterative process; an optimal iteration count controls the balance between achieving a satisfactory solution and maintaining reasonable computational resources. The algorithm's pseudocode is delineated in Algorithm~\ref{alg:pulse_sequence_optimization}.

\begin{algorithm}
\color{black}
\caption{Pulse Optimization Algorithm}
\label{alg:pulse_sequence_optimization}
\begin{algorithmic}[1]
\Require \(\Omega_{\text{bounds}}\), \(\delta_{\text{bounds}}\), \(T_{\text{bounds}}\) \Comment{Parameter intervals for Rabi frequency, detuning, and pulse duration.}
\Require \(Register\), \(H_P\), \(p\) \Comment{Quantum register, Problem Hamiltonian, Number of iterations \(p\).}
\State Initialize \(params\) as the initial solution (the average of parameter bounds).
\For{\(i = 1\) to \(p\)}
    \State \(cost, samples \gets\) EvaluateSequence(\(params\))
    \If{\(cost <\) best cost found so far}
        \State Update \(best\_params\) and \(best\_samples\) with current \(params\) and \(samples\).
    \EndIf
\EndFor
    \State \(params \gets\) Generate new parameters based on GBRT optimization or initial parameters for the first iteration.

\State \Return \(best\_params\), \(best samples\).
\Function{EvaluateSequence}{params}
    \State Generate the pulse sequence with \(params\).
    \State Apply the sequence to the quantum system.
    \State Measure the outcome to collect samples.
    \State Calculate the average cost $\langle C \rangle$.
    \State \Return the average cost $\langle C \rangle$, samples.
\EndFunction
\end{algorithmic}
\end{algorithm}

\begin{figure}[!h]
\color{black}
    \centering
    \includegraphics[width=9cm,height=3.7cm]{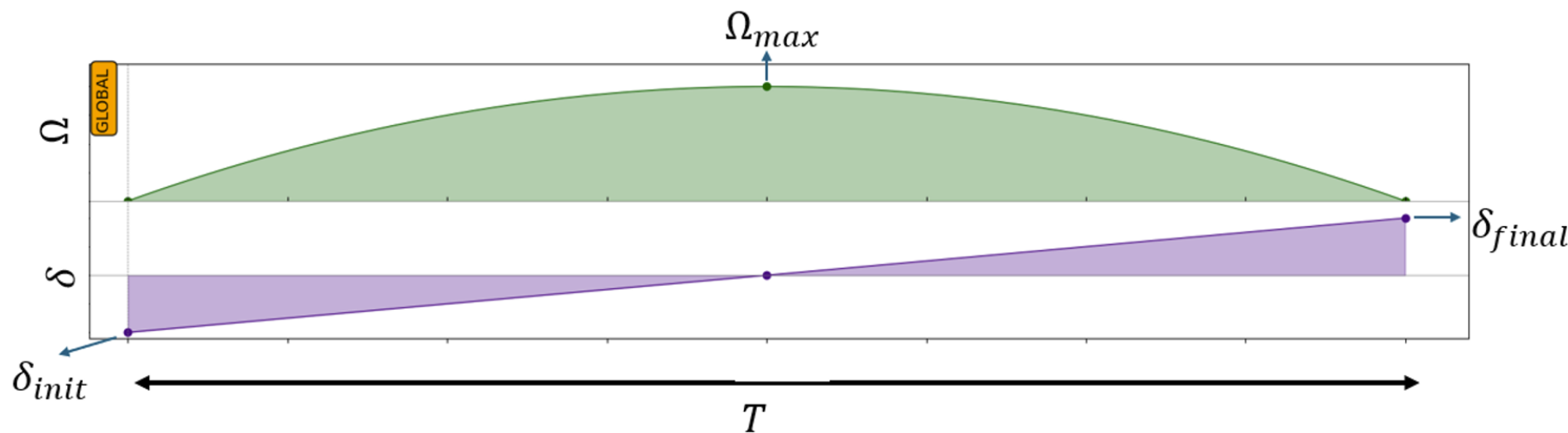}
    \caption{Shape of an Optimized Pulse. The figure demonstrates the change in Rabi frequency (\(\Omega\)) and global detuning (\(\delta\)) during the pulse  (\(T\)). \(\Omega\) exhibits a peak at \(\Omega_{\text{max}}\) while \(\delta\) varies linearly from \(\delta_{\text{init}}\) to \(\delta_{\text{final}}\).}
    \label{fig:adiabatic_pulse}
\end{figure}

Figure~\ref{fig:adiabatic_pulse} shows an example of an optimized pulse corresponding to a gradual parameter evolution over an extended duration \(T\). This is particularly used by the Quantum Adiabatic Algorithm~\cite{albash2018adiabatic}, which emerges as a potent strategy for efficiently addressing the optimization problem.

\color{black}
\section{The Hybrid Quantum-Classical Benders Algorithm}
\label{section:hybridBenders}

\begin{figure}[!htb]
\centering
\includegraphics[width=1\linewidth]{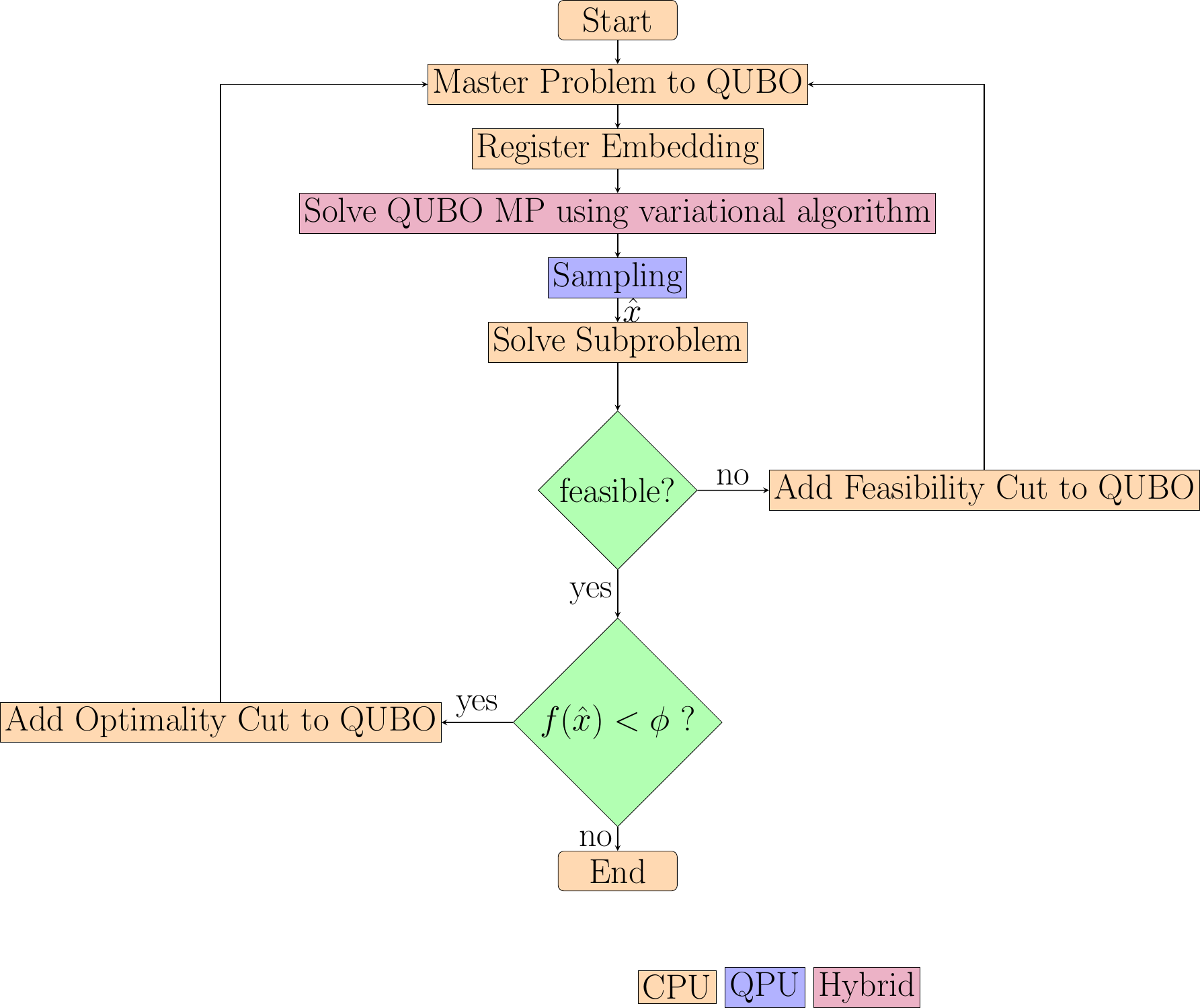} 
\caption{Hybrid BD Quantum-Classical Algorithm Flowchart.
The flowchart presents the sequence of steps in our hybrid quantum-classical algorithm. Starting with the reformulation of the MP into a QUBO, the process employs Register Embedding to spatially arrange atoms for QUBO encoding. This is followed by the application of \textcolor{black}{the variational algorithm}, where pulse parameters are optimized for quantum sampling. The QPU then samples solutions, with the best candidate \(\hat{x}\) selected based on the costs. if the SP is infeasible, a feasibility cut is introduced into the MP. If feasible, an optimality cut is applied if necessary (\(f(\hat{x}) < \phi) \)). The process iterates between these steps until the best solution is found, resulting in the final values of \(x\) and \(y\).}
\label{fig:hybrid_algorithm}
\end{figure}

This section presents the overall BD Hybrid Quantum-Classical Algorithm, which uses elements discussed in Section~\ref{section:background} and the study from Sections~\ref{section:to_qubo} and~\ref{section:quantumalgo}. The algorithm combines quantum and classical computing resources to address MILP problems using the BD algorithm. The sequence of operations and decision-making processes involved in this algorithm is represented in the flowchart of Figure~\ref{fig:hybrid_algorithm}.

The algorithm uses the decomposition of the OP, as introduced in \eqref{OP} -  \eqref{y-variables}, into a MP and a SP. Initially, the MP  \eqref{master-problem} - \eqref{master:y} is reformulated into a Quadratic Unconstrained Binary Optimization (QUBO) model \eqref{QUBO}. Subsequently, the Register Embedding heuristic, detailed in Algorithm~\ref{alg:register_embedding}, is employed. This heuristic arranges atoms in a spatial configuration that closely aligns the interaction matrix with the QUBO.

Following the register embedding, the \textcolor{black}{variational algorithm} is applied. In this step, pulse parameters are tuned according to the established register configuration. Next, the QPU is used as a sampler. Multiple measurements are performed on the final quantum states. These measurements yield various potential solutions for the OP, each with an associated probability of occurrence and cost. The solution $\hat{x}$, which yields the lowest cost, is selected for further processing.

The next phase involves classical computing methods to solve the SP \eqref{sp:objective} - \eqref{sp:variables}. The feasibility of its solution is assessed. If the SP is infeasible, a feasibility Benders' cut \eqref{feasibility-cut} is generated and added into the MP, redirecting the process to the QUBO reformulation. Conversely, if the solution is feasible, a subsequent check is conducted to compare the optimal objective value of the SP \(f(\hat{x})\) with the value of the variable \(\phi\). In case \(f(\hat{x})\) be lower, an optimality cut \eqref{optimality-cut} is added to the MP, and the algorithm revisits the QUBO reformulation step. If not, the process stops, outputing the latest values of $x$ and $y$ as the solution obtained by the Hybrid BD algorithm.

\begin{table*}[t]
\centering
\begin{tabular}{|p{3cm}|p{7cm}|p{7cm}|}
\hline
 & \textbf{\textcolor{black}{Iteration 1}} & \textbf{\textcolor{black}{Iteration 2}} \\
\hline
\textbf{\textcolor{black}{MP QUBO}} & 
\textcolor{black}{$-(15 x_1^2 + 10 x_2^2) + 100 (-x_1^2 -x_2^2 + 2^0 s_{m\_1\_1} + 1)^2$ \newline
$+ (2^4 w_1^2 + 2^3 w_2^2 + 2^2 w_3^2 + 2^1 w_4^2 + 2^0 w_5^2)$} & \textcolor{black}{MP(it1)+ generated penalty} \\
\hline
\textbf{\textcolor{black}{MP solution}} & 
\textcolor{black}{$\Phi: 31.5$ \newline 
$obj: 21.5$ \newline 
$x_1: 0,$ $x_2: 1,$ \newline 
$s_{m\_1\_1}: 0$} & 
\textcolor{black}{$\Phi: 17.0$ \newline 
$obj: 2.0$ \newline 
$x_1: 1,$ $x_2: 0,$ \newline 
$s_{m\_1\_1}: 0,$ $s1\_1: 0$} \\
\hline
\textbf{\textcolor{black}{SP solution}} & 
\textcolor{black}{Objective Value: 11.0\newline 
$y_1: 0.0,$ $y_2: 0.0,$ $y_3: 1.0,$ $y_4: 1.0$ \newline 
$\mu_1: 5.0,$ $\mu_2: 0.0,$ $\mu_3: 0.0,$ $\mu_4: 6.0,$ \newline 
$\mu_5: -3.0,$ $\mu_6: -3.0,$ $\mu_7: 0.0,$ $\mu_8: 0.0$} & 
\textcolor{black}{Objective Value: 17.0\newline 
$y_1: 1.0,$ $y_2: 1.0,$ $y_3: 0.0,$ $y_4: 0.0$ \newline 
$\mu_1: 8.0,$ $\mu_2: 0.0,$ $\mu_3: 0.0,$ $\mu_4: 9.0,$ \newline 
$\mu_5: 0.0,$ $\mu_6: 0.0,$ $\mu_7: 0.0,$ $\mu_8: 0.0$} \\
\hline
\textbf{\textcolor{black}{Generated penalty}} & 
\textcolor{black}{$100 \left(2^4 w_1 + 2^3 w_2 + 2^2 w_3 \right.$ \newline
$\left. + 2^1 w_4 + 2^0 w_5 + 3.0 \times (-x_1) + 3.0 \times (-x_2) \right.$ \newline
$\left. + 2^0 s1\_1 - (5.0 + 6.0)\right)^2$} & \textcolor{black}{--} \\
\hline
\textbf{\textcolor{black}{Type of penalty}} & \textcolor{black}{Optimality} & \textcolor{black}{--} \\
\hline
\end{tabular}
\caption{\textcolor{black}{Trace of the BD algorithm solving process on the MILP}}
\label{poc}
\end{table*}

\section{Numerical experiments}
\label{section:experimentation}

In this section, we present our numerical experiments. The primary objective is to provide a PoC. Moreover, on a set of arbitrary MILP small instances, we conduct a comparatif study between our hybrid BD solver and classical BD, with the MP solved using simulated annealing. 

\subsection{Proof of Concept (PoC)}
For the PoC, we use the same example presented in~\cite{zhao2022hybrid}. The MILP associated with the problem uses two binary variables $ x_1, x_2 \in \{0, 1\}$, and four non-negative continuous variables $ y_1, y_2, y_3, y_4\geq 0$. The matrix description of the MILP is:


\[
A = \begin{bmatrix}
0 & 0 \\
0 & 0 \\
0 & 0 \\
0 & 0 \\
-1 & 0 \\
-1 & 0 \\
0 & -1 \\
0 & -1
\end{bmatrix}, 
\quad
G = \begin{bmatrix}
1 & 0 & 1 & 0 \\
1 & 0 & 0 & 1 \\
0 & 1 & 1 & 0 \\
0 & 1 & 0 & 1 \\
1 & 0 & 0 & 0 \\
0 & 1 & 0 & 0 \\
0 & 0 & 1 & 0 \\
0 & 0 & 0 & 1
\end{bmatrix}, 
\quad
b = \begin{bmatrix}
1 \\
1 \\
1 \\
1 \\
0 \\
0 \\
0 \\
0
\end{bmatrix}, 
\]
\[
B = \begin{bmatrix}
-1 & -1 \\
\end{bmatrix}, 
\quad
b' = \begin{bmatrix}
-1 \\
\end{bmatrix}, 
\]
\[
c^T = \begin{bmatrix} -15 & -10 \end{bmatrix}, 
\quad
 h^T = \begin{bmatrix} 8 & 9 & 5 & 6 \end{bmatrix}, 
\]

In the presented use case, we set the penaty coefficients to: penalty of the master objective function \( \pi_{\text{obj}} = 1 \) and penalties of the constraints of the master constraints as well as optimality and feasibility Benders' constraints, respectively to \( \pi_1 = \pi_2 = \pi_3 = 100 \). Table~\ref{poc} outlines the execution state of the BD algorithm applied to the example described above. The table includes:

\begin{itemize}
    \item \textbf{MP QUBO:} The literal expression of the initial MP.
    \item \textbf{MP solution:} The solutions obtained from the quantum sampler: values of variables \( x \), \( s \), \( \phi \), and the objective function value, denoted as $obj$.
    \item \textbf{SP solution:} The solution to the SP: objective value, values of variables \( y \) solution, and its dual \( \mu \).
    \item \textbf{Generated penalty:} Literal expression of the penalty (if any).
    \item \textbf{Type of penalty:} Type of the penalty (if any).
\end{itemize}


\color{black}
Solving the use case using the hybrid BD algorithm, the final objective value is equal to the optimal solution optimal objective value determined by CPLEX.
Moreover, as shown in Table~\ref{poc}, our hybrid BD algorithm with neutral atoms concludes within two iterations, which marks an improvement compared to the five iterations required by the hybrid Bender's algorithm presented in~\cite{zhao2022hybrid}, using D-Wave.
This faster convergence can be attributed to the fact that our quantum sampler help in providing stronger Benders' cuts. The neutral atom QPU generates better MP solutions. Using these solutions, the SP yield tighter values of the left hand side of optimality cuts~\eqref{optimality-cut}. Consequently, the variable \( \phi \) converge more rapidly, allowing to confirm optimality in fewer iteration.

The distinctive advantages of using neutral atoms in our BD algorithm are primarily grounded in the superior problem encoding capabilities of neutral atom QPUs. In D-Wave systems, the embedding procedure involves linking chains of qubits with a strongly ferromagnetic coupling (\(K=-2\)) to simulate a single logical variable. Additionally, these logical variables are then coupled with \(Q_{ij}\), which is constrained within the narrow range \([-1,1]\) (for D-Wave, \(Q_{ij} = J_{ij}\))~\cite{dwaveembedding}. This restrictive range necessitates rescaling all couplings to fit within it, which can compromise the fidelity of the problem representation.

By contrast, our approach with neutral atoms leverages long-range Rydberg interactions, allowing for a much more faithful embedding. The Rydberg dipole-dipole interactions, characterized by \(C_6/r^6\), provide a mechanism to finely tune the interaction distances. This ability enables a versatile adjustment of interaction strengths over a broad spectrum due to the power-law decay, offering significant advantages in problem encoding.
For our 9-variable all-to-all connected QUBO, this means we can handle the MILP problem's extensively varying coupling values without the need to rescale them, thus preserving the fidelity of the problem representation. This is particularly beneficial as our QUBO model features an exponentially broad range of coupling values, which poses a challenge in D-Wave systems where lower \(J_{ij}\) values can be drowned in thermal noise. Furthermore, the architecture of neutral atoms facilitates coherent annealing processes, where the dephasing and depolarizing timescales exceed the annealing period. This ensures that the system's evolution is predominantly isolated from the thermal environment, enhancing computational performance and stability.

The enhanced performance of our algorithm is not only attributable to the quantum hardware employed but also significantly influenced by our pre-processing strategy, which optimizes algorithm convergence. As detailed in Section~\ref{section:to_qubo}, by solving linear relaxations and integrating the OP valid constraints, our strategy tightens the upper bounds of \(\phi_{\text{max}}\) and the slack variables \(s_{\text{max}}^k\).  By providing a tighter upper bound, we help the algorithm reach the optimal solution more quickly because the tighter bounds limit the solution space the algorithm needs to explore, thus accelerating convergence.
Additionally, this approach reduces the quantum resource requirements by decreasing the number of qubits needed for encoding, thereby enhancing resource utilization.

\color{black}
This use-case demonstrates the potential of neutral atoms QPUs when integrated into a hybrid BD algorithm; it  confirms the feasibility of applying such quantum computational resources and shows their ability to enhance performance, outperforming current state-of-the-art solutions.

\subsection{Numerical results}
We examine now the efficiency of the hybrid BD algorithm, on several MILPs instances.  
\subsubsection{MILPs description and implementation features}

The instances generation process of our experimentation covers 450 MILPs randomly generated. We vary the number of variables and constraints, as well as the structure and coefficients of the constraint matrices as the following:

\begin{itemize}
    \item \textbf{Variables \( x \) and \( y \)}: The number of binary variables \( x \) range from 2 to 5, while the number continuous variables \( y \) vary from 2 to 10. This setting is motivating by the exploration of different problem scales, while respecting the capacity of the simulation in terms of number of qubits.
    \item \textbf{Constraint Matrices \( A \), \( G \) and Vector \( b \)}: Matrix \( A \) is populated with non-positive random values, and matrix \( G \) with non-negative values. This setting is choosen to yield positive slack variables, thereby reducing the number of required qubits needed in converting optimality and feasibility cuts to the QUBO formulation. The vector $b$ is randomly generated with non-negative random values.  These parameters define the set of constraints \eqref{linking-constraints}, whose number vary from 5 to 14. 
    \item \textbf{Matrix \( B \) and Vector \( b' \)}: Matrix \( B \) consists of a single row filled with ones, and vector \( b' \) is a random positive number strictly less than the size of \( x \). This imposes a special constraint that interdicts choosing the solution where all the $x$ variables are equal to $1$. These parameters are related to constraints \eqref{x-constraints}. Only one constraint of this type is considered.  
    \item \textbf{Coefficients \( c \) and \( h \)}: These are randomly generated vectors with non-negative random values, contributing to the variability in the objective function.
\end{itemize}

The randomness in these parameters generates various instances.  We group the results based on the number of qubits used during the whole process, and average the output for each qubit count.
The evaluation of the algorithm performance is based on three key metrics:

\begin{itemize}
    \item \textbf{Percentage of Feasible Instances}: This metric evaluates the algorithm's ability to find feasible solutions for the tested instances.
    \item \textbf{Gap to Optimality}: This metric indicates the quality of the solution obtained by the algorithm. It is defined as \( \text{Gap} = \frac{\text{obj(algo)} - \text{obj(opt)}}{\text{obj(opt)}} \), where obj is the objective function value. The Gap represents the distance of the solution provided by the algorithm from the optimal solution. The optimal solution is computed considering the OP in its compact formulation before being reformulated for a BD.
    \item \textbf{Number of Iterations}: This metric serves as an indicator of the time and energy consumption of the algorithm.
\end{itemize}

For the computational implementation and analysis of the generated MILP instances, the programming is conducted in Python. The quantum pulses are simulated using Pulser~\cite{PulserDocs}, which is used for designing and emulating quantum protocols on neutral atom devices. The \textcolor{black}{variational algorithm} resolution is conducted by scikit-optimize~\cite{scikit-optimizeDocs}, a Python library for optimization that is well-suited for quantum algorithm parameter tuning. Finally, the compact formulation of the OP was solved to optimality using the CPLEX solver~\cite{IBMCPLEX2023}, a high-performance mathematical programming solver.

\subsubsection{Results}

We benchmark our algorithm with a fully classical BD where we solve the MP using simulated annealing. It is important to note that our tests limits the number of qubits to 11, which is a threshold set by the simulation constraints of current quantum simulator capabilities. 

\begin{figure}[!htb]
\color{black}
    \centering
    \includegraphics[width=0.85\linewidth]{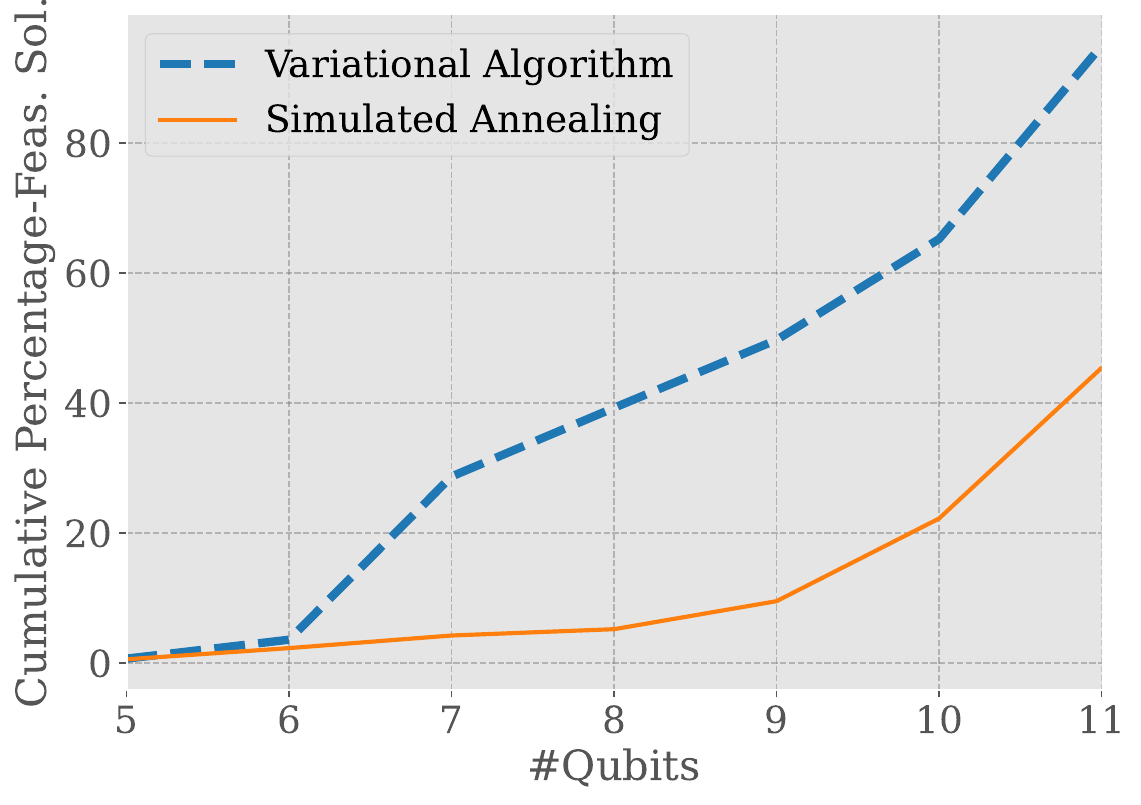}
    \caption{Percentage of feasible solutions per number of qubits}
    \label{fig:qaoa_vs_sa}
\end{figure}

The graphic in Figure~\ref{fig:qaoa_vs_sa} illustrates the comparative performance of our hybrid BD with \textcolor{black}{the variational algorithm} and classical BD using simulated annealing, in terms of the cumulative percentage of feasible solutions, with respect to the number of qubits. It is evident that \textcolor{black}{the variational algorithm} surpasses simulated annealing consistently throughout the observed qubit range. \textcolor{black}{Our variational algorithm} demonstrates a pronounced increase in the cumulative percentage of feasible solutions, as the qubit count increases, achieving more than $95 \%$ with 11 qubits. In contrast, simulated annealing shows a more moderate 
progression.

\begin{figure}[!htb]
\color{black}
    \centering
    \includegraphics[width=1\linewidth]{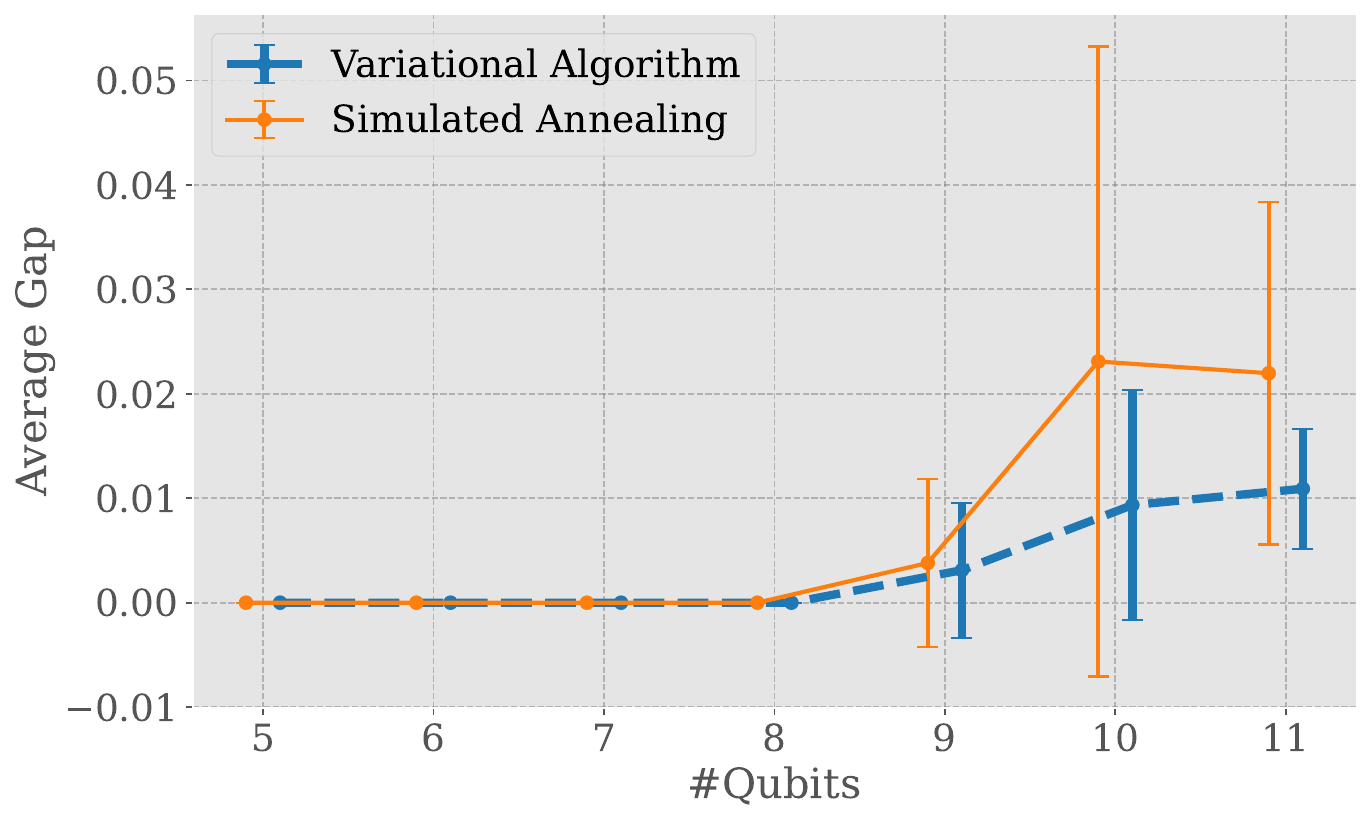}
    \caption{Average gap per number of qubits}
    \label{fig:average_gap}
\end{figure}

Figure~\ref{fig:average_gap} shows the average gap to optimality (and
the standard deviation with 95\% confidence interval), for both hybrid BD with \textcolor{black}{the variational algorithm} and fully classical BD with Simulated Annealing in function of  the number of qubits. The gap to optimality serves as a crucial measure of solution quality. A smaller gap indicative of a solution that is nearer to the optimal.
The results presented here were conducted on MILPs where both methods, \textcolor{black}{variational algorithm} and Simulated Annealing, provided feasible solutions (thus, on the 45\% of feasible MILPs given by Simulated Annealing). This ensures the fact that the gap is defined for both \textcolor{black}{variational algorithm} and Simulated Annealing. It can be seen that overall, both methods deliver solutions of good quality. The maximum average gap is attributed to Simulated Annealing and is equal to 2.3\%. Nonetheless, 
observations indicate that \textcolor{black}{the variational algorithm} maintains a relatively stable average gap, suggesting a robust ability to generate solutions close to the optimal across different qubit counts. Conversely, Simulated Annealing exhibits a comparable performance at lower qubit counts but deteriorates as the count becomes greater than 10. This performance degradation shows the  potential scalability issues with Simulated Annealing when faced with increased problem complexity, represented by higher qubit counts, and consequently the potential ability of our algorithm to produce a high quality solutions when scaling-up.

\begin{figure}[!htb]
\color{black}
    \centering
    \includegraphics[width=1\linewidth]{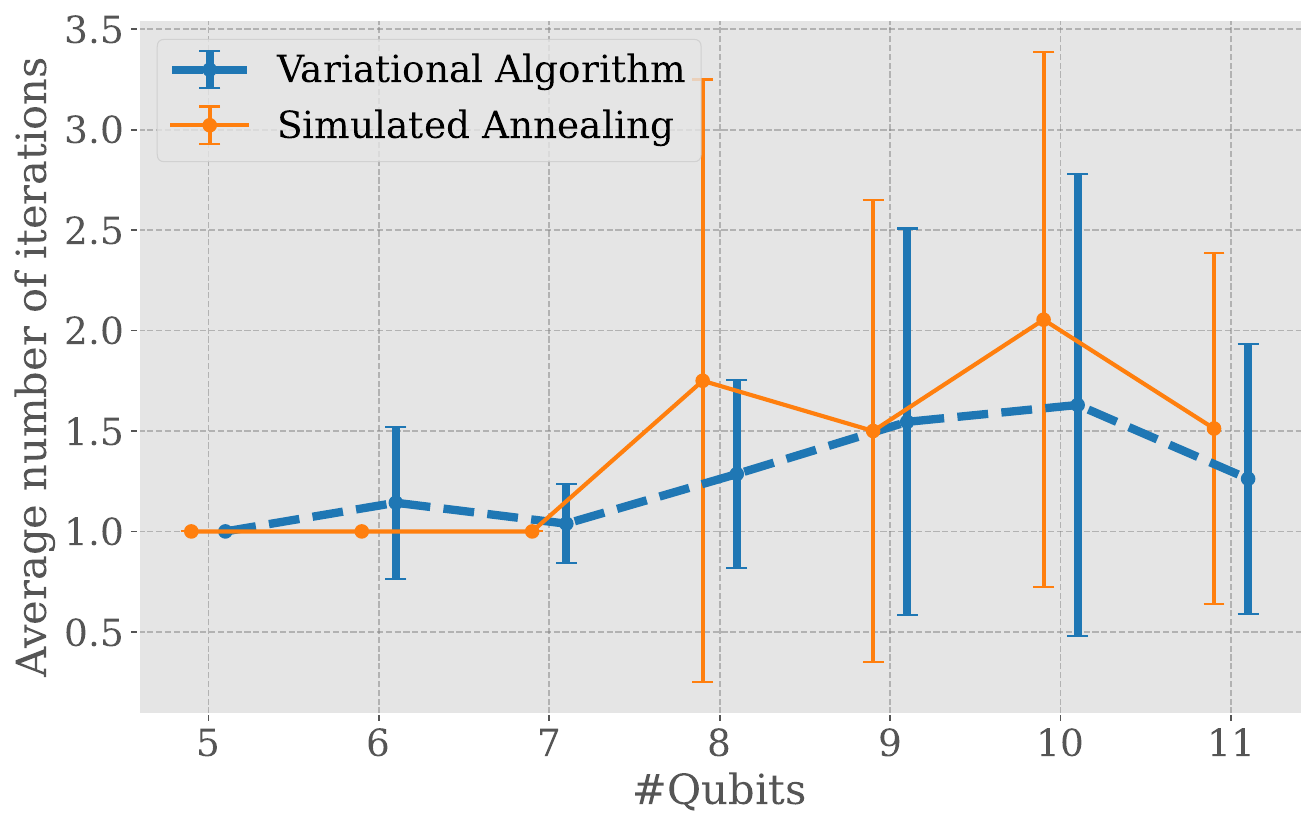}
    \caption{Average number of BD iterations per number of qubits}
    \label{fig:iterations_qaoa_sa}
\end{figure}

Figure~\ref{fig:iterations_qaoa_sa} presents the average number of iterations (and
the standard deviation with 95\% confidence interval), for the hybrid BD with \textcolor{black}{variational algorithm} and the fully classical BD with Simulated Annealing as a function of the number of qubits. Iterations reflect the computational effort and, by extension, time and energy expenditure of the algorithms. Both algorithms show an increase in iterations with more qubits. This is explained by the complexity generated by the number of qubits. Simulated Annealing count increases at 10 qubits, indicating possible inefficiencies at this problem size. In contrast, \textcolor{black}{the variational algorithm} displays a moderate increase, showing a more stable scaling performance. This is particularly remarkable at 11 qubits where the number of iterations decreases. 

In conclusion, the numerical results, on this type of MILPs, show the efficiency of our hybrid BD approach in comparison to the fully classical Benders method with simulated annealing. The results affirm that the hybrid BD  generates a higher percentage of feasible solutions for various qubit counts, and maintains a closer distance to optimal solutions. Moreover, the hybrid BD algorithm shows a more stable performance with respect to the number of iterations, especially with a larger number of qubits. These results show the potential of hybrid BD algorithm with neutral atoms to efficiently handle larger and more complex problems, even within the current limitations of quantum computational resources. 
Future work will aim to extend these investigations beyond the 11-qubit threshold as advancements in quantum computing technology become available. It is also important to note that extending these preliminary tests to more complex MILP scenarios may reveal instances where the classical BD with simulated annealing outperforms the hybrid BD with \textcolor{black}{the variational algorithm}, due to qubit limitations and/or algorithmic convergence issues. In such cases, a further research study would be conducted. This will be discussed in Section \ref{section:conclusion}.

\section{Conclusion and perspectives}
\label{section:conclusion}

In this study, we assisted classical BD with neutral atoms computation, to address MILP problems. To this end, we have designed a hybrid classical-quantum algorithm. We have developed an automated procedure to transform the MP into a QUBO formulation. We also presented a heuristic for register embedding. In addition, we implemented a \textcolor{black}{variational algorithm} for pulse shaping. A PoC has shown our method applicability, and preliminary numerical results has demonstrated the efficiency of our hybrid framework which outperforms fully classical BD techniques. While this research shows the potential of hybrid quantum-classical BD algorithms, supported by neutral atom computation, in addressing MILPs, it opens new avenues for future advancements in this area.
It should be noted that while the results presented in this paper are encouraging, they apply only to a specific set of small instances. Expanding these findings and extending the scale of application are not included in the scope of this initial PoC work.

Looking to the future, our objective is to scale and diversify our instances. We are aware of the potential challenges of this step, especially in terms of qubit resource limitations and algorithmic convergence. In case we encounter resource limitations in terms of number of qubits, one promising avenue is to set a maximum number of qubits as a computational threshold and implement an iterative process that aims to guarantee the quality of the solution, while respecting the qubit count limitation. More precisely, within each iteration, one can evaluate the current penalty terms and their contribution to the solution quality. Any penalty term not significantly influencing the solution can be removed to free up computational resources. This allows for the introduction of new penalty terms that may further refine the solution. This process should be repeated until the solution converges. By managing the number of qubits in this manner, we hope to achieve a balance between qubit utilization and solution quality. This futurist approach potentially enables us to solve complex problems within a limited quantum environment.

In order to address the potential convergence issues of the algorithm, we can consider multiple MP solutions, in each iteration. To this end, we can build on the interesting work of~\cite{paterakis2023hybrid}, which is based on the concepts of \textit{multi-cuts} introduced by~\cite{beheshti2019accelerating}. This method involves generating multiple solutions for the MP and then selecting a specific subset of Benders' cuts to not surcharge the MP. By solving a set covering problem, we can identify the minimal set of constraints necessary to exclude all suboptimal or infeasible MP solutions. This approach can speed up the algorithm convergence and, at the same time, reduce the number of qubits needed. Moreover, the task of identifying the optimal subset of constraints is equivalent to solving a MIS problem, making it well-suited for execution on a neutral atoms QPU.


\section*{Acknowlegments}
We thank Constantin Dalyac for insightful discussions, as well as Anna Joliot for her helpful outcome in some experimentation. 

\bibliographystyle{IEEEtran}
\bibliography{bib}

\begin{thebibliography}{10}
\providecommand{\url}[1]{#1}
\csname url@samestyle\endcsname
\providecommand{\newblock}{\relax}
\providecommand{\bibinfo}[2]{#2}
\providecommand{\BIBentrySTDinterwordspacing}{\spaceskip=0pt\relax}
\providecommand{\BIBentryALTinterwordstretchfactor}{4}
\providecommand{\BIBentryALTinterwordspacing}{\spaceskip=\fontdimen2\font plus
\BIBentryALTinterwordstretchfactor\fontdimen3\font minus \fontdimen4\font\relax}
\providecommand{\BIBforeignlanguage}[2]{{%
\expandafter\ifx\csname l@#1\endcsname\relax
\typeout{** WARNING: IEEEtran.bst: No hyphenation pattern has been}%
\typeout{** loaded for the language `#1'. Using the pattern for}%
\typeout{** the default language instead.}%
\else
\language=\csname l@#1\endcsname
\fi
#2}}
\providecommand{\BIBdecl}{\relax}
\BIBdecl

\bibitem{paschos2014applications}
V.~T. Paschos, \emph{Applications of combinatorial optimization}.\hskip 1em plus 0.5em minus 0.4em\relax John Wiley \& Sons, 2014, vol.~3.

\bibitem{benichou1971experiments}
M.~B{\'e}nichou, J.-M. Gauthier, P.~Girodet, G.~Hentges, G.~Ribi{\`e}re, and O.~Vincent, ``Experiments in mixed-integer linear programming,'' \emph{Mathematical Programming}, vol.~1, pp. 76--94, 1971.

\bibitem{vanderbei2020linear}
R.~J. Vanderbei \emph{et~al.}, \emph{Linear programming}.\hskip 1em plus 0.5em minus 0.4em\relax Springer, 2020.

\bibitem{marchand2002cutting}
H.~Marchand, A.~Martin, R.~Weismantel, and L.~Wolsey, ``Cutting planes in integer and mixed integer programming,'' \emph{Discrete Applied Mathematics}, vol. 123, no. 1-3, pp. 397--446, 2002.

\bibitem{bnnobrs1962partitioning}
J.~BnnoBRs, ``Partitioning procedures for solving mixed-variables programming problems,'' \emph{Numer. Math}, vol.~4, no.~1, pp. 238--252, 1962.

\bibitem{vanderbeck2006generic}
F.~Vanderbeck and M.~W. Savelsbergh, ``A generic view of dantzig--wolfe decomposition in mixed integer programming,'' \emph{Operations Research Letters}, vol.~34, no.~3, pp. 296--306, 2006.

\bibitem{rahmaniani2017Benders}
R.~Rahmaniani, T.~G. Crainic, M.~Gendreau, and W.~Rei, ``The benders decomposition algorithm: A literature review,'' \emph{European Journal of Operational Research}, vol. 259, no.~3, pp. 801--817, 2017.

\bibitem{farhi2014quantum}
E.~Farhi, J.~Goldstone, and S.~Gutmann, ``A quantum approximate optimization algorithm,'' \emph{arXiv preprint arXiv:1411.4028}, 2014.

\bibitem{mcclean2016theory}
J.~R. McClean, J.~Romero, R.~Babbush, and A.~Aspuru-Guzik, ``The theory of variational hybrid quantum-classical algorithms,'' \emph{New Journal of Physics}, vol.~18, no.~2, p. 023023, 2016.

\bibitem{nannicini2019performance}
G.~Nannicini, ``Performance of hybrid quantum-classical variational heuristics for combinatorial optimization,'' \emph{Physical Review E}, vol.~99, no.~1, p. 013304, 2019.

\bibitem{li2017hybrid}
J.~Li, X.~Yang, X.~Peng, and C.-P. Sun, ``Hybrid quantum-classical approach to quantum optimal control,'' \emph{Physical review letters}, vol. 118, no.~15, p. 150503, 2017.

\bibitem{da2023quantum}
W.~da~Silva~Coelho, L.~Henriet, and L.-P. Henry, ``Quantum pricing-based column-generation framework for hard combinatorial problems,'' \emph{Physical Review A}, vol. 107, no.~3, p. 032426, 2023.

\bibitem{zhao2022hybrid}
Z.~Zhao, L.~Fan, and Z.~Han, ``Hybrid quantum benders’ decomposition for mixed-integer linear programming,'' in \emph{2022 IEEE Wireless Communications and Networking Conference (WCNC)}.\hskip 1em plus 0.5em minus 0.4em\relax IEEE, 2022, pp. 2536--2540.

\bibitem{gao2022hybrid}
F.~Gao, D.~Huang, Z.~Zhao, W.~Dai, M.~Yang, and F.~Shuang, ``Hybrid quantum-classical general benders decomposition algorithm for unit commitment with multiple networked microgrids,'' \emph{arXiv preprint arXiv:2210.06678}, 2022.

\bibitem{chang2020hybrid}
C.-Y. Chang, E.~Jones, Y.~Yao, P.~Graf, and R.~Jain, ``On hybrid quantum and classical computing algorithms for mixed-integer programming,'' \emph{arXiv preprint arXiv:2010.07852}, 2020.

\bibitem{fan2022hybrid}
L.~Fan and Z.~Han, ``Hybrid quantum-classical computing for future network optimization,'' \emph{IEEE Network}, vol.~36, no.~5, pp. 72--76, 2022.

\bibitem{nguyen2023quantum}
M.-T. Nguyen, J.-G. Liu, J.~Wurtz, M.~D. Lukin, S.-T. Wang, and H.~Pichler, ``Quantum optimization with arbitrary connectivity using rydberg atom arrays,'' \emph{PRX Quantum}, vol.~4, no.~1, p. 010316, 2023.

\bibitem{kjaergaard2020superconducting}
M.~Kjaergaard, M.~E. Schwartz, J.~Braum{\"u}ller, P.~Krantz, J.~I.-J. Wang, S.~Gustavsson, and W.~D. Oliver, ``Superconducting qubits: Current state of play,'' \emph{Annual Review of Condensed Matter Physics}, vol.~11, pp. 369--395, 2020.

\bibitem{pogorelov2021compact}
I.~Pogorelov, T.~Feldker, C.~D. Marciniak, L.~Postler, G.~Jacob, O.~Krieglsteiner, V.~Podlesnic, M.~Meth, V.~Negnevitsky, M.~Stadler \emph{et~al.}, ``Compact ion-trap quantum computing demonstrator,'' \emph{PRX Quantum}, vol.~2, no.~2, p. 020343, 2021.

\bibitem{anton2019interfacing}
C.~Ant{\'o}n, J.~C. Loredo, G.~Coppola, H.~Ollivier, N.~Viggianiello, A.~Harouri, N.~Somaschi, A.~Crespi, I.~Sagnes, A.~Lemaitre \emph{et~al.}, ``Interfacing scalable photonic platforms: solid-state based multi-photon interference in a reconfigurable glass chip,'' \emph{Optica}, vol.~6, no.~12, pp. 1471--1477, 2019.

\bibitem{glover2018tutorial}
F.~Glover, G.~Kochenberger, and Y.~Du, ``A tutorial on formulating and using qubo models,'' \emph{arXiv preprint arXiv:1811.11538}, 2018.

\bibitem{IBMCPLEX2023}
\BIBentryALTinterwordspacing
IBM, ``Cplex optimizer,'' 2023. [Online]. Available: \url{https://www.ibm.com/fr-fr/analytics/cplex-optimizer}
\BIBentrySTDinterwordspacing

\bibitem{GurobiOptimization}
\BIBentryALTinterwordspacing
{Gurobi Optimization}, ``Gurobi - the fastest solver - gurobi,'' 2023, accessed: 2023-12-15. [Online]. Available: \url{https://www.gurobi.com}
\BIBentrySTDinterwordspacing

\bibitem{bertsimas1993simulated}
D.~Bertsimas and J.~Tsitsiklis, ``Simulated annealing,'' \emph{Statistical science}, vol.~8, no.~1, pp. 10--15, 1993.

\bibitem{desaulniers2006column}
G.~Desaulniers, J.~Desrosiers, and M.~M. Solomon, \emph{Column generation}.\hskip 1em plus 0.5em minus 0.4em\relax Springer Science \& Business Media, 2006, vol.~5.

\bibitem{do2023quantum}
D.~T. Do, N.~Trieu, and D.~T. Nguyen, ``Quantum-based distributed algorithms for edge node placement and workload allocation,'' \emph{arXiv preprint arXiv:2306.01159}, 2023.

\bibitem{wille2019ibm}
R.~Wille, R.~Van~Meter, and Y.~Naveh, ``Ibm’s qiskit tool chain: Working with and developing for real quantum computers,'' in \emph{2019 Design, Automation \& Test in Europe Conference \& Exhibition (DATE)}.\hskip 1em plus 0.5em minus 0.4em\relax IEEE, 2019, pp. 1234--1240.

\bibitem{boyd2011distributed}
S.~Boyd, N.~Parikh, E.~Chu, B.~Peleato, J.~Eckstein \emph{et~al.}, ``Distributed optimization and statistical learning via the alternating direction method of multipliers,'' \emph{Foundations and Trends{\textregistered} in Machine learning}, vol.~3, no.~1, pp. 1--122, 2011.

\bibitem{naoum2013interior}
J.~Naoum-Sawaya and S.~Elhedhli, ``An interior-point benders based branch-and-cut algorithm for mixed integer programs,'' \emph{Annals of Operations Research}, vol. 210, pp. 33--55, 2013.

\bibitem{crainic2014partial}
T.~G. Crainic, M.~Hewitt, and W.~Rei, \emph{Partial decomposition strategies for two-stage stochastic integer programs}.\hskip 1em plus 0.5em minus 0.4em\relax CIRRELT, 2014, vol.~88.

\bibitem{yang2012tighter}
Y.~Yang and J.~M. Lee, ``A tighter cut generation strategy for acceleration of benders decomposition,'' \emph{Computers \& Chemical Engineering}, vol.~44, pp. 84--93, 2012.

\bibitem{fabian2007solving}
C.~I. F{\'a}bi{\'a}n and Z.~Sz{\H{o}}ke, ``Solving two-stage stochastic programming problems with level decomposition,'' \emph{Computational Management Science}, vol.~4, pp. 313--353, 2007.

\bibitem{rubiales2013stabilization}
A.~J. Rubiales, P.~A. Lotito, and L.~A. Parente, ``Stabilization of the generalized benders decomposition applied to short-term hydrothermal coordination problem,'' \emph{IEEE Latin America Transactions}, vol.~11, no.~5, pp. 1212--1224, 2013.

\bibitem{beheshti2019accelerating}
N.~Beheshti~Asl and S.~MirHassani, ``Accelerating benders decomposition: multiple cuts via multiple solutions,'' \emph{Journal of Combinatorial Optimization}, vol.~37, pp. 806--826, 2019.

\bibitem{franco2023efficient}
N.~Franco, T.~Wollschl{\"a}ger, B.~Poggel, S.~G{\"u}nnemann, and J.~M. Lorenz, ``Efficient milp decomposition in quantum computing for relu network robustness,'' \emph{arXiv preprint arXiv:2305.00472}, 2023.

\bibitem{tarjan1977finding}
R.~E. Tarjan and A.~E. Trojanowski, ``Finding a maximum independent set,'' \emph{SIAM Journal on Computing}, vol.~6, no.~3, pp. 537--546, 1977.

\bibitem{clark1990unit}
B.~N. Clark, C.~J. Colbourn, and D.~S. Johnson, ``Unit disk graphs,'' \emph{Discrete mathematics}, vol.~86, no. 1-3, pp. 165--177, 1990.

\bibitem{stastny2023functional}
S.~Stastny, H.~P. B{\"u}chler, and N.~Lang, ``Functional completeness of planar rydberg blockade structures,'' \emph{Physical Review B}, vol. 108, no.~8, p. 085138, 2023.

\bibitem{Sakurai_ModernQM}
J.~J. Sakurai and J.~Napolitano, \emph{Modern Quantum Mechanics}, 2nd~ed.\hskip 1em plus 0.5em minus 0.4em\relax Addison-Wesley, 2011, chapter 5 provides a detailed discussion on the adiabatic theorem in quantum mechanics.

\bibitem{da2022efficient}
W.~da~Silva~Coelho, M.~D’Arcangelo, and L.-P. Henry, ``Efficient protocol for solving combinatorial graph problems on neutral-atom quantum processors,'' \emph{arXiv preprint arXiv:2207.13030}, 2022.

\bibitem{novo2018environment}
L.~Novo, S.~Chakraborty, M.~Mohseni, and Y.~Omar, ``Environment-assisted analog quantum search,'' \emph{Physical Review A}, vol.~98, no.~2, p. 022316, 2018.

\bibitem{henriet2020quantum}
L.~Henriet, L.~Beguin, A.~Signoles, T.~Lahaye, A.~Browaeys, G.-O. Reymond, and C.~Jurczak, ``Quantum computing with neutral atoms,'' \emph{Quantum}, vol.~4, p. 327, 2020.

\bibitem{Benders2005partitioning}
J.~F. Benders, ``Partitioning procedures for solving mixed-variables programming problems,'' \emph{Computational Management Science}, vol.~2, no.~1, pp. 3--19, 2005.

\bibitem{balinski1969duality}
M.~Balinski and A.~W. Tucker, ``Duality theory of linear programs: A constructive approach with applications,'' \emph{Siam Review}, vol.~11, no.~3, pp. 347--377, 1969.

\bibitem{schrijver1998theory}
A.~Schrijver, \emph{Theory of linear and integer programming}.\hskip 1em plus 0.5em minus 0.4em\relax John Wiley \& Sons, 1998.

\bibitem{bertsimas1997introduction}
D.~Bertsimas and J.~N. Tsitsiklis, \emph{Introduction to linear optimization}.\hskip 1em plus 0.5em minus 0.4em\relax Athena scientific Belmont, MA, 1997, vol.~6.

\bibitem{ayodele2022penalty}
M.~Ayodele, ``Penalty weights in qubo formulations: Permutation problems,'' in \emph{European Conference on Evolutionary Computation in Combinatorial Optimization (Part of EvoStar)}.\hskip 1em plus 0.5em minus 0.4em\relax Springer, 2022, pp. 159--174.

\bibitem{verma2022penalty}
A.~Verma and M.~Lewis, ``Penalty and partitioning techniques to improve performance of qubo solvers,'' \emph{Discrete Optimization}, vol.~44, p. 100594, 2022.

\bibitem{garcia2022exact}
M.~D. Garc{\'\i}a, M.~Ayodele, and A.~Moraglio, ``Exact and sequential penalty weights in quadratic unconstrained binary optimisation with a digital annealer,'' in \emph{Proceedings of the Genetic and Evolutionary Computation Conference Companion}, 2022, pp. 184--187.

\bibitem{prettenhofer2014gradient}
P.~Prettenhofer and G.~Louppe, ``Gradient boosted regression trees in scikit-learn,'' in \emph{PyData 2014}, 2014.

\bibitem{albash2018adiabatic}
T.~Albash and D.~A. Lidar, ``Adiabatic quantum computation,'' \emph{Reviews of Modern Physics}, vol.~90, no.~1, p. 015002, 2018.

\bibitem{dwaveembedding}
``D-wave embedding,'' \url{https://docs.ocean.dwavesys.com/en/stable/concepts/embedding.html}, D-Wave Systems.

\bibitem{PulserDocs}
\BIBentryALTinterwordspacing
{Pulser Development Team}, ``Pulser documentation.'' [Online]. Available: \url{https://pulser.readthedocs.io/en/stable/}
\BIBentrySTDinterwordspacing

\bibitem{scikit-optimizeDocs}
\BIBentryALTinterwordspacing
{scikit-optimize Developers}, ``scikit-optimize: sequential model-based optimization in python,'' Online Documentation. [Online]. Available: \url{https://scikit-optimize.github.io/stable/}
\BIBentrySTDinterwordspacing

\bibitem{paterakis2023hybrid}
N.~G. Paterakis, ``Hybrid quantum-classical multi-cut benders approach with a power system application,'' \emph{Computers \& Chemical Engineering}, vol. 172, p. 108161, 2023.

\end{thebibliography}

\end{document}